\documentclass{article}
\usepackage[a4paper, left=1.2in, right=1.2in, top=1in, bottom=1in]{geometry}

\usepackage{booktabs}
\usepackage{graphicx} 
\usepackage{xcolor}
\usepackage{amsmath}
\usepackage{amssymb}
\usepackage{natbib}
\usepackage{hyperref}
\usepackage{comment}
\usepackage{tikz}
\usepackage{tabularx}
\usepackage{subcaption}
\usetikzlibrary{matrix}
\usepackage{xstring} 
\usetikzlibrary{arrows.meta, positioning}
 
\title{Multiplex relational event models of self-reported and perceived gossip}

\author{
Veronica Poda$^{a}$\thanks{ \texttt{veronica.poda@unitn.it}},\ 
Veronica Vinciotti$^{a}$\thanks{\texttt{veronica.vinciotti@unitn.it}},\ 
Ernst C.\ Wit$^{b}$\thanks{\texttt{ernst.jan.camiel.wit@usi.ch}}
}

\date{}

\begin{document}

\maketitle

\vspace{-0.5cm}

\begin{center}
$^{a}$Department of Mathematics, University of Trento, Trento, Italy\\
$^{b}$Institute of Computing, Università della Svizzera italiana, Lugano, Switzerland
\end{center}

\vspace{0.5cm}

\begin{abstract}
\noindent Gossiping is a complex social phenomenon that unfolds through multiple types of relations. Among other things, individuals may initiate gossiping about others as a result of believing themselves to be the target of gossip. These two processes, i.e., gossiping and gossip perception, form a dynamic multiplex network, where self-reported and perceived gossip ties coexist and can influence each other over time. We develop a multiplex relational event model to study the joint evolution of these two processes. 

The proposed model includes endogeneous covariates that capture the influence of events occurring within one layer, as well as events occurring on the other layer,  on the rate of self-reported and perceived gossiping. We analyze survey data collected annually from 44 secondary schools in Hungary. We show how a multiplex relational event model, including linear, smooth and random effects, is able to describe the social mechanisms operating within and between self-reported and perceived gossiping and to capture how these two layers influence each other over time. 

\vspace{0.2cm}

\end{abstract}

\section{Introduction} \label{sec:introduction}

Gossiping consists of interactions among individuals who exchange information and opinions about others. Such interactions can be of a positive nature, and result in the strengthening of social bonds, but also of a negative nature, such as when they contribute to bullying \citep{bullying, bullies}. The general aim of this paper is to describe the underlying dynamics of gossiping. To this end, different dimensions of gossiping are important. Individual \textit{gossiping} about another person may be affected, among other things, by the gossiper \textit{perceiving} that gossiping is directed towards them by that person. This means that the two processes are interdependent, with perceived gossiping influencing subsequent gossiping behaviour, and vice versa. In order to account for these different processes, and have a more complete characterization of the gossiping dynamics, this paper considers a multiplex representation of gossiping, where different types of gossiping can influence each other over time.

Several approaches have been developed for modelling multiplex networks, both in the case of static and dynamic networks \citep{wang2013ergmextensions, chen2021statistical, snijders2013model}. These approaches are primarily formulated in terms of network ties or network states, rather than time-ordered relational events. The latter is instead the case of gossiping, where events happen instantaneously in continuous time. Motivated by this, \cite{poda2025hypereventnetworkmodellingpartially} develop a relational event model of self-reported gossiping. However, they do not account for the effect of perceived gossiping, or, in general, of other types of gossiping that may effect self-reporting gossiping.    
 \citet{nynkegossip}, on the other hand,  study specifically the dependence between self-reported and perceived gossiping. However, the multiplex $p_2$ model that they develop is only suitable for static gossiping data, while information on time ordering plays a key role in understanding the dynamics of gossiping. Moreover, their analysis focuses primarily on whether individuals accurately perceive gossiping directed towards them, rather than on the role of perceived gossiping in influencing future gossiping behaviour and vice versa.  

To address these limitations, in this paper, we propose a multiplex Relational Event Model (REM). This is an extension of standard REMs for jointly modelling time-ordered events across multiple relational layers. 
In the specific case of gossiping, this model allows to account for self-reported and perceived gossiping as two interdependent relational event processes, allowing past interactions in one layer to affect subsequent interactions in the other. We illustrate the proposed framework with an empirical application to longitudinal school survey data collected annually from  a number of secondary school classes in Hungary  \citep{gossip_dataset}. Two questions in the survey specifically focus on self-reported and perceived gossiping, allowing us to study their interconnected dynamics.

We define endogenous covariates that explicitly account for the multiplex nature of the relations. In this way, the model accounts for dependences both within each event type and between different event types.  We then develop statistical inference for multiplex REMs on rather generic relational data structures. In particular, we start from the availability of complete relational event data and then move to the case of aggregated counts across time intervals. Finally, we also consider the case of right-censored interval-time data. This is particularly relevant for the empirical illustration, where for each layer we only know whether at least one gossiping event of that type occurred during the time period covered by each survey. In all these cases, we show how efficient generalized additive models can be used to fit multiplex REMs. A combination of linear, smooth and random effects can therefore be included to capture the complexity of the underlying process and quantify the effects of important drivers. 


The remainder of this paper is organized as follows. Section~\ref{sec:motivating_example} motivates the methodological development using the school survey data from \cite{gossip_dataset}. We then introduce the multiplex REM in section~\ref{sec:multiplexREM}, including the definition of endogenous covariates across the different multiplex layers. Section~\ref{sec:inference} presents inferential approaches for complete, aggregated and censored relational event data. Section~\ref{sec:simulation} evaluates the proposed method through simulation studies, showing the advantages of a multiplex formulation of the problem compared to using standard REMs on each separate layer with no cross-layer dependences. 
Finally, section~\ref{sec:realdata} goes back to the motivating example and studies the dynamics of perceived and self-reported gossiping via a suitably defined multiplex REM, with  complex within-layer and between-layer linear and smooth effects.

\section{RECENS: a school survey study involving  gossiping}
\label{sec:motivating_example}

We study the dynamics of gossiping within and between different dimensions using longitudinal survey data from the \textit{Wired into Each Other} project of the Research Center for Educational and Network Studies (RECENS) in Hungary \citep{gossip_dataset}. The study tracks $1,695$ students across $44$ classes in $7$ secondary schools over $4$ annual survey waves. Data collection began two months after the start of high school (October-November 2010), targeting 9th-grade students with an average age of 15.1 years. Subsequent waves were conducted six months later (April 2011), one year after the second wave (April 2012), and finally one year after the third wave (April 2013) when the students were in 12th grade. Among the various survey questions, we focus on two in particular which measure self-reported and perceived gossip, respectively. The first question, \textit{Of whom do you say bad things to your friends?}, captures self-reported gossip and relates to the actual student's behaviour. Here, we use a directed tie $s \to r$ to indicate that student $s$ reports gossiping about student $r$. The second question, \textit{Who says bad things about you?}, captures perceived gossip, that is how student's behaviour is perceived by others. In this case, we use a directed tie $s \to r$ to indicate that student $r$ perceives student $s$ as gossiping about them.

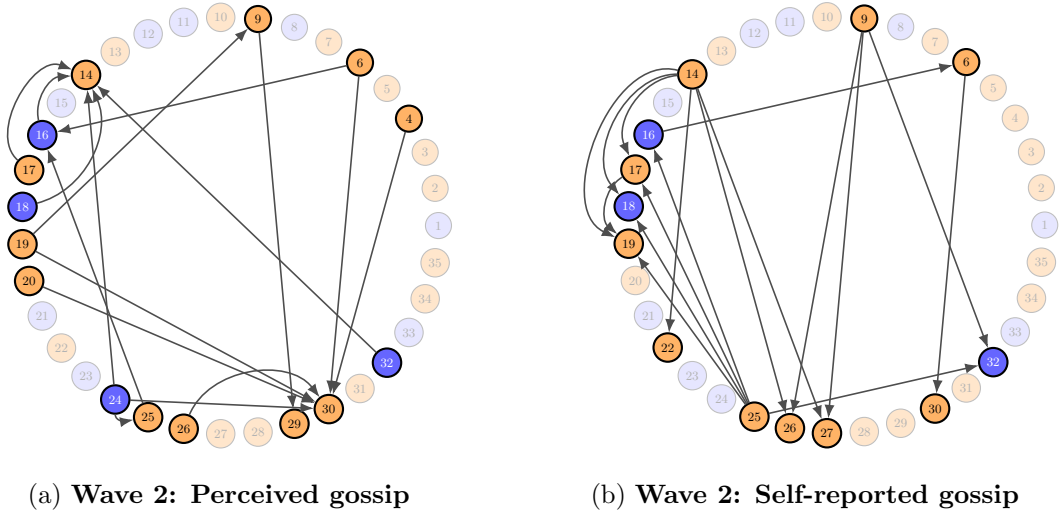
\begin{figure}[t]
\tikzset{>={Latex[scale=0.9]}} 
\centering
\def\allnodes{1401,1402,1403,1404,1405,1406,1407,1408,1409,1410,1411,1412,1413,1414,1415,1416,1417,1418,1419,1420,1421,1422,1423,1424,1425,1426,1427,1428,1429,1430,1431,1432,1433,1434,1435}

\def\genderMap{1401/0,1402/1,1403/1,1404/1,1405/1,1406/1,1407/1,1408/0,1409/1,1410/1,1411/0,1412/0,1413/1,1414/1,1415/0,1416/0,1417/1,1418/0,1419/1,1420/1,1421/0,1422/1,1423/0,1424/0,1425/1,1426/1,1427/1,1428/1,1429/1,1430/1,1431/1,1432/0,1433/0,1434/1,1435/1}

\begin{minipage}[t]{0.48\textwidth}
\centering
\begin{tikzpicture}[scale=0.53, transform shape]
    \pgfmathtruncatemacro{\numAllNodes}{35}
    \foreach [count=\i] \num in \allnodes {
        \pgfmathsetmacro{\angle}{360/\numAllNodes*(\i-1)}
        \coordinate (pos\num) at (\angle:5.2cm);
    }

    \def\activeNodes{1404,1406,1409,1414,1416,1417,1418,1419,1420,1424,1425,1426,1429,1430,1432}
    
    \foreach \num/\gen in \genderMap {
        \pgfmathtruncatemacro{\labelnum}{mod(\num,100)}
        \def\isActive{0}
        \foreach \act in \activeNodes {
            \ifnum\num=\act \gdef\isActive{1} \fi
        }
        
        \ifnum\isActive=1
            \ifnum\gen=0 
                \def\colorfill{blue!60!white}\def\colortext{white}
            \else
                \def\colorfill{orange!60}\def\colortext{black}
            \fi
            \node[circle, draw=black, thick, fill=\colorfill, text=\colortext, minimum size=6.5mm] (n\num) at (pos\num) {\small\labelnum};
        \else
            \ifnum\gen=0
                \def\colorfillfade{blue!10!white}
            \else                \def\colorfillfade{orange!20!white}
            \fi
            \node[circle, draw=black!25, fill=\colorfillfade, minimum size=6.5mm, text=black!30] (n\num) at (pos\num) {\small\labelnum};
        \fi
    }

    \draw[black!70, semithick, ->] (n1404) -- (n1430);
    \draw[black!70, semithick, ->] (n1406) -- (n1416);
    \draw[black!70, semithick, ->] (n1406) -- (n1430);
    \draw[black!70, semithick, ->] (n1409) -- (n1429);
    \draw[black!70, semithick, ->] (n1419) -- (n1409);
    \draw[black!70, semithick, ->] (n1419) -- (n1430);
    \draw[black!70, semithick, ->] (n1420) -- (n1430);
    \draw[black!70, semithick, ->] (n1424) -- (n1414);
    \draw[black!70, semithick, ->] (n1424) -- (n1430);
    \draw[black!70, semithick, ->] (n1425) -- (n1416);
    \draw[black!70, semithick, ->] (n1432) -- (n1414);
    \draw[black!70, semithick, ->] (n1418) to[bend right=50] (n1414);
    \draw[black!70, semithick, ->] (n1417) to[bend left=80] (n1414);
    \draw[black!70, semithick, ->] (n1416) to[bend left=50] (n1414);
    \draw[black!70, semithick, ->] (n1424) to[out=270, in=190] (n1425);
    \draw[black!70, semithick, ->] (n1426) to[bend left=60] (n1430);
\end{tikzpicture}

\vspace{0.3cm}
(a) \textbf{Wave 2: Perceived gossip}
\end{minipage}%
\hfill
\begin{minipage}[t]{0.48\textwidth}
\centering
\begin{tikzpicture}[scale=0.53, transform shape]
    \pgfmathtruncatemacro{\numAllNodes}{35}
    \foreach [count=\i] \num in \allnodes {
        \pgfmathsetmacro{\angle}{360/\numAllNodes*(\i-1)}
        \coordinate (pos\num) at (\angle:5.2cm);
    }

    \def\activeNodes{1406,1409,1414,1416,1417,1418,1419,1422,1425,1426,1427,1430,1432}
    
    \foreach \num/\gen in \genderMap {
        \pgfmathtruncatemacro{\labelnum}{mod(\num,100)}
        \def\isActive{0}
        \foreach \act in \activeNodes {
            \ifnum\num=\act \gdef\isActive{1} \fi
        }
        
        \ifnum\isActive=1
            \ifnum\gen=0 
                \def\colorfill{blue!60!white}\def\colortext{white}
            \else
                \def\colorfill{orange!60}\def\colortext{black}
            \fi
            \node[circle, draw=black, thick, fill=\colorfill, text=\colortext, minimum size=6.5mm] (n\num) at (pos\num) {\small\labelnum};
        \else
            \ifnum\gen=0
                \def\colorfillfade{blue!10!white}
            \else
                \def\colorfillfade{orange!20!white}
            \fi
            \node[circle, draw=black!25, fill=\colorfillfade, minimum size=6.5mm, text=black!30] (n\num) at (pos\num) {\small\labelnum};
        \fi
    }

    \draw[black!70, semithick, ->] (n1416) -- (n1406);
    \draw[black!70, semithick, ->] (n1425) -- (n1416);
    \draw[black!70, semithick, ->] (n1425) -- (n1417);
    \draw[black!70, semithick, ->] (n1425) -- (n1418);
    \draw[black!70, semithick, ->] (n1425) -- (n1419);
    \draw[black!70, semithick, ->] (n1414) -- (n1422);
    \draw[black!70, semithick, ->] (n1409) -- (n1426);
    \draw[black!70, semithick, ->] (n1414) -- (n1426);
    \draw[black!70, semithick, ->] (n1409) -- (n1427);
    \draw[black!70, semithick, ->] (n1414) -- (n1427);
    \draw[black!70, semithick, ->] (n1406) -- (n1430);
    \draw[black!70, semithick, ->] (n1409) -- (n1432);
    \draw[black!70, semithick, ->] (n1425) -- (n1432);

    \draw[black!70, semithick, ->] (n1414) to[bend right=55] (n1417);
    \draw[black!70, semithick, ->] (n1414) to[bend right=65] (n1418);
    \draw[black!70, semithick, ->] (n1414) to[bend right=85] (n1419);
    \draw[black!70, semithick, ->] (n1417) to[bend right=55] (n1419);
\end{tikzpicture}

\vspace{0.3cm}
(b) \textbf{Wave 2: Self-reported gossip}
\end{minipage}

\caption{For a representative class of $35$ students from the school survey data, (a) perceived gossip ties and  (b) self-reported gossip ties at wave $2$ of the survey are visualized.  Female students are represented in orange and male students in blue. Inactive nodes are shown with faded colors corresponding to their gender. The fact that student 16 gossips about student 6 may be retaliation from the belief that student 6 is gossiping about him.}
\label{fig:multiplex_gosnet}
\end{figure}

As discussed in section~\ref{sec:introduction}, we hypothesize that the two processes influence each other.
Figure~\ref{fig:multiplex_gosnet} visualizes the data for one representative classroom of $35$ students and one wave. 
Figure~\ref{fig:multiplex_gosnet}a shows the collection of perceived gossip ties, while Figure~\ref{fig:multiplex_gosnet}b shows the self-reported gossip ties. It is remarkable how different the two networks are, meaning that accurate perception of actual gossip tends to be quite poor. This suggests that the generative process of both processes is quite distinct, but not necessarily unrelated. For example, student 16 may be gossiping about student 6 in wave 2, because he, falsely, believed that she was gossiping about him. This cross-network retaliation also seems to be at play for student 14 gossiping about student 17. A joint modelling approach of the two processes allows one to detect the presence, and quantify the extent of, such a cross-layer effect from the full time-ordered sequence of events of the two different types. 

\begin{figure}[t]
    \centering
    \hfill
    \begin{minipage}[b]{0.48\textwidth}
    \centering
\resizebox{!}{1.8cm}{
        \begin{tabular}{lr}
        \toprule
        \textbf{Characteristic} & \textbf{Value} \\
        \midrule
        Students & 1,695 \\
        Waves & 4 \\
        Schools & 7 \\
        Classes & 44 \\
        Self-reported gossip events & 3,482 \\
        Perceived gossip events & 2,183 \\
        Risk set size & 143,130 \\
        \bottomrule
        \end{tabular}}  
        
   \vspace{0.8cm}
        (a)
    \end{minipage}
       \hfill 
    \begin{minipage}[b]{0.48\textwidth}
        \centering
        \includegraphics[width=\linewidth]{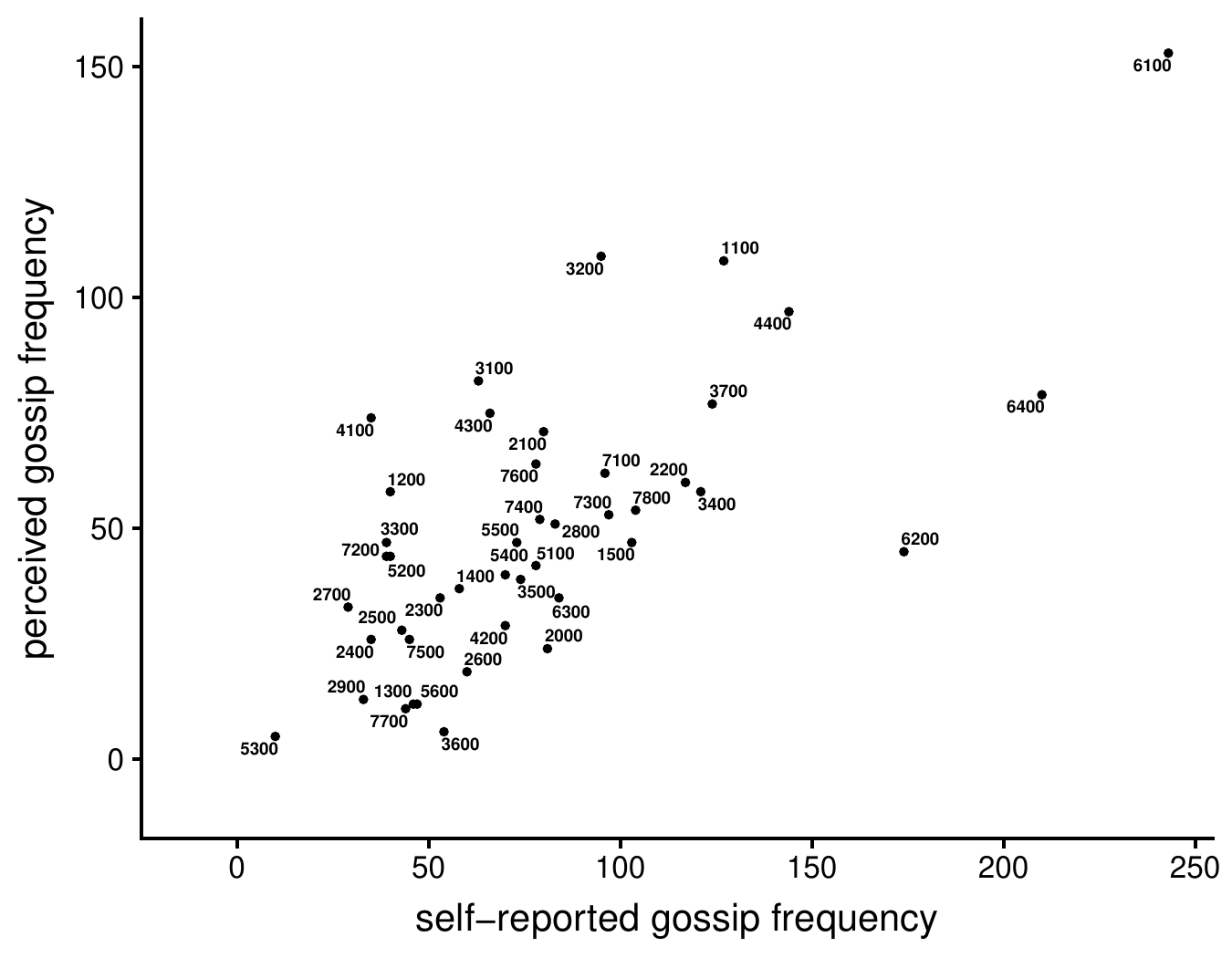}
        \\
        (b)
    \end{minipage}
    \hfill
    \begin{minipage}[b]{0.48\textwidth}
        \centering
        \includegraphics[width=\linewidth]{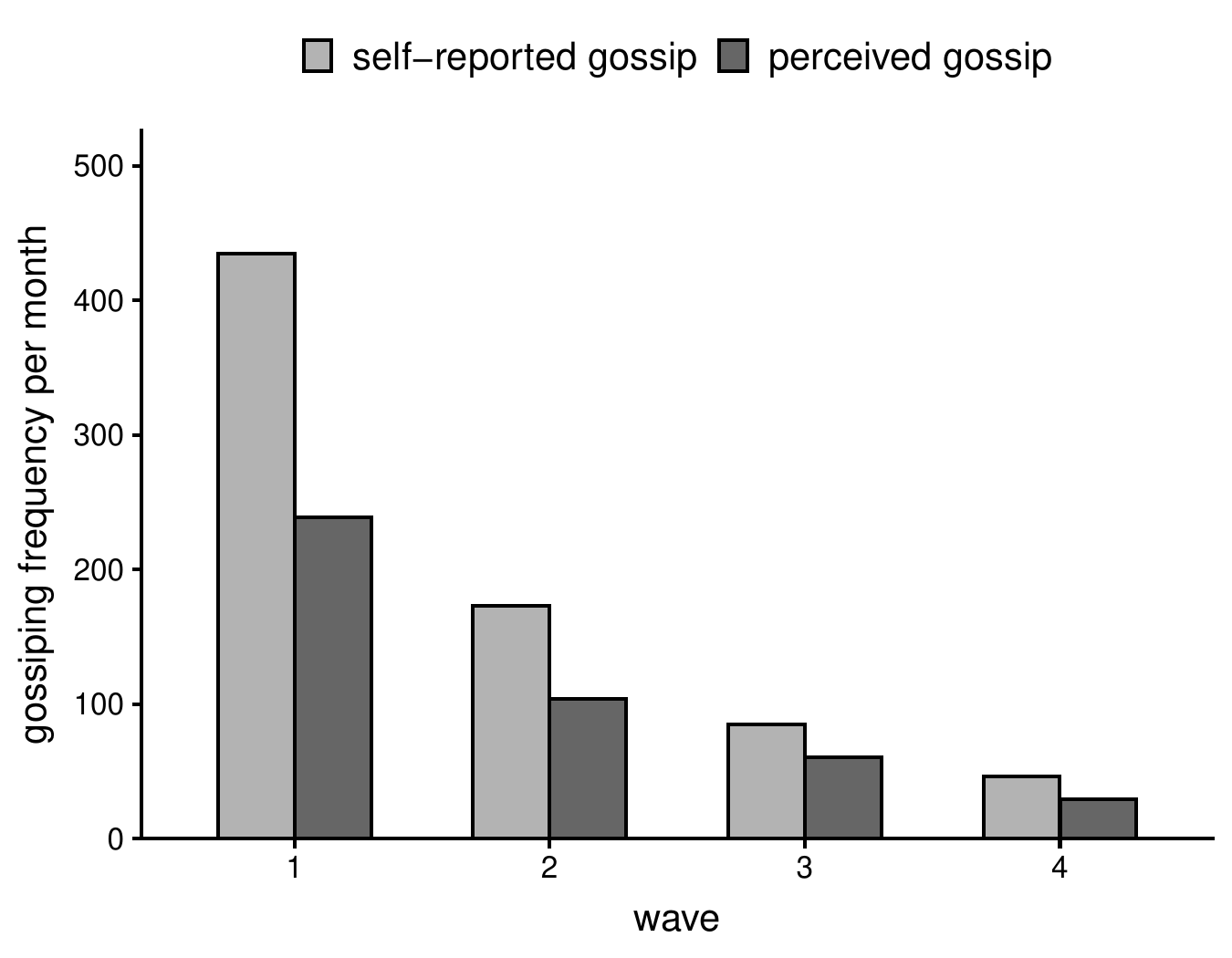}\\
        (c)
    \end{minipage}
       \hfill    
    \begin{minipage}[b]{0.48\textwidth}
        \centering
        \includegraphics[width=\linewidth]{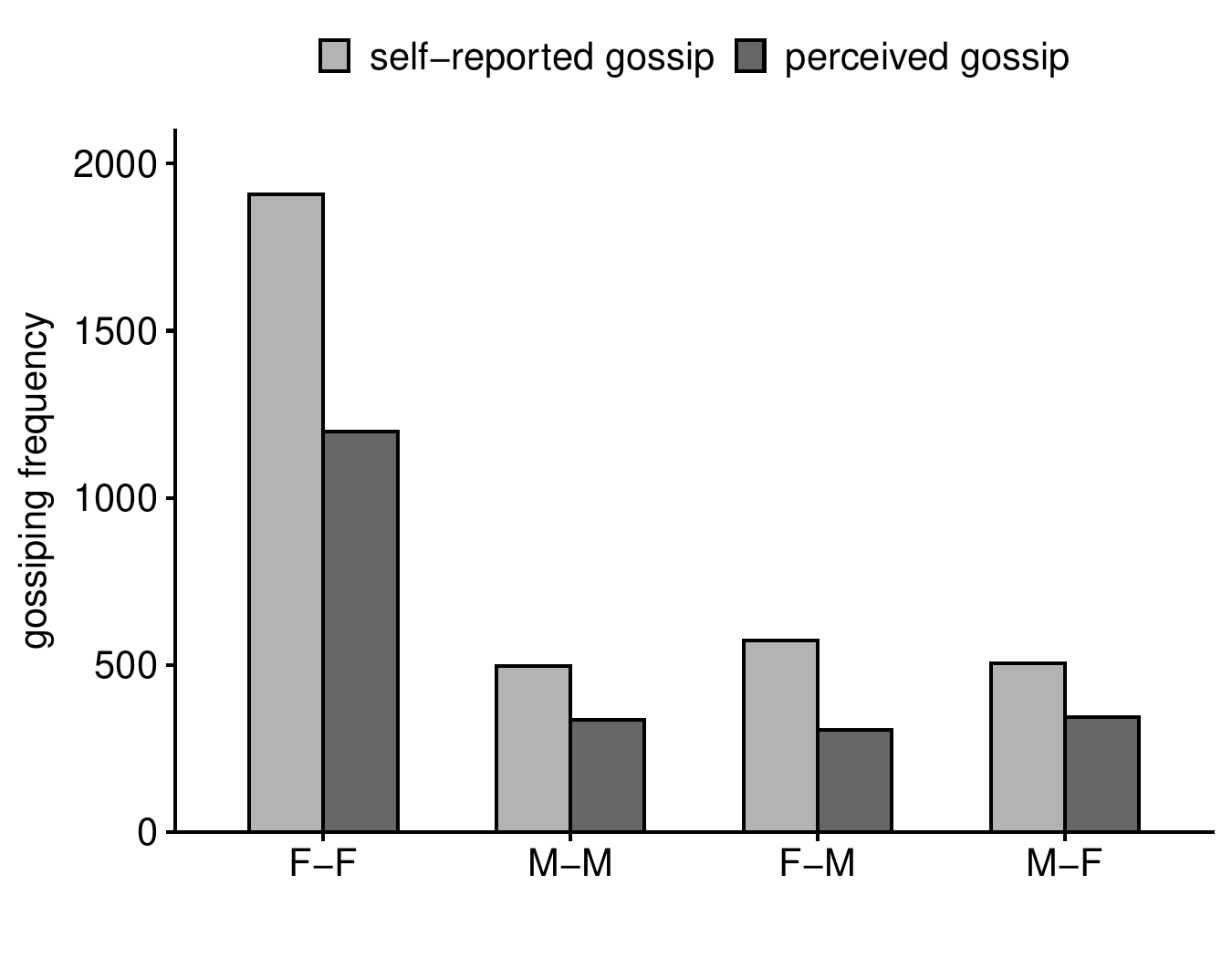}
        \\
        (d)
    \end{minipage}

    \caption{(a) Summary statistics of the school survey data. (b) Self-reported and perceived gossiping are correlated and show heterogeneity across classes. (c) Self-reported and perceived gossiping decrease over time. (d) Female to female gossiping accounts for the highest proportion of gossiping events in both networks. }
    \label{fig:gossip_figures}
\end{figure}

Figures~\ref{fig:gossip_figures} provides summaries of the full data across all waves and classrooms. In total, we observe $3,482$ self-reported gossiping events and $2,183$ perceived gossiping events, out of a risk set of $143,130$ possible directed ties (Figures~\ref{fig:gossip_figures}a). Figures~\ref{fig:gossip_figures}c shows a decrease in gossiping over time, both for the self-reported and perceived gossiping, while Figures~\ref{fig:gossip_figures}d shows how female to female gossiping accounts for the highest proportion of gossiping events in both networks.   Figures~\ref{fig:gossip_figures}b shows  a large heterogeneity across classes: some classes display a high level of gossiping, whereas others remain relatively quiet. This tends to be consistent across the self-reported and perceived gossiping, possibly as a result of a dependence between the two processes.  For example, class 6100 is the most active in both self-reported and perceived gossiping, whereas class 5300 is the least active in both. However, this is not the case for all classes, e.g., class 3100 exhibits relatively high levels of perceived gossiping but low levels of self-reported gossiping. This shows how other covariates may also play a role in the co-evolution of the two processes.

The multiplex relational event model that we develop in this paper will allow us to study the interdependence between the self-reported and perceived gossiping as well as the effect of other covariates on gossiping. Random effects can also be included in the model at each layer, e.g., sender, receiver and class-specific in the case of the school survey data, in order to account for additional heterogeneity in the underlying process which is not accounted for by the covariates.


 %

\section{Multiplex relational event model}
\label{sec:multiplexREM}
We now describe the multiplex relational event framework. Section~\ref{sec:model} formalizes the underling multiplex relational event process and thus provides the generic formulation of the model. In the following section~\ref{sec:multiplex_statistics}, we describe a number of covariates that capture effects across layers in the co-evolution of the processes.

\subsection{Model formulation} \label{sec:model}

In a standard relational event model \citep{bianchi2023}, an event is defined by a sender, a receiver, and the time it occurs. A multiplex relational event extends this framework by additionally assigning an event type to each interaction. Formally, we define a multiplex event as the tuple
\begin{equation*}
(s, r, q, t),
\end{equation*}
where $s$ is the sender, $r$ the receiver, $q$ the event type, and $t\in [0,T]$ the time of the event. Here, $s \in V_S(t)$, $r \in V_R(t)$, and $q \in Q$, where $V_S(t)$ and $V_R(t)$ are the sets of potential senders and receivers at time $t$, respectively, and $Q$ is the set of event types. Importantly, the set of actors is the same across the different event types and the same sender-receiver pair can be involved in multiple event types, motivating a multiplex view of the relational event process.
In the gossiping illustration of section~\ref{sec:motivating_example}, there are two event types, namely the self-reported and perceived gossiping. These form two interacting layers of the multiplex relational event network.
An event in one layer may be followed by an event in the other, suggesting that past events in one layer may influence the occurrence of subsequent events in the other. This motivates a joint model formulation, in which the history of both layers can contribute to the intensity of each event type.

In order to formalize this, the multiplex events are assumed to be generated by a marked point process
\begin{equation*}
\{(s_i,r_i,q_i,t_i) \, | \, i \geq 1, \, t_i \in [0,T], \, s_i \in V_S(t_i), \, r_i \in V_R(t_i),\, q_i \in Q\}.
\end{equation*}
At each event time $t_i$, an interaction occurs between a sender $s_i$ and a receiver $r_i$ with event type $q_i$. We associate a multivariate counting process $N$, which counts the number of directed interactions from $s$ to $r$ of type $q$ up to time $t$:
\begin{equation*}
N_{srq}(t) = \sum_{i \geq 1} 1\{t_i \leq t,\ s_i = s,\ r_i = r, \ q_i = q\}.
\end{equation*}
By the Doob–Meyer decomposition \citep{meyer1962decomposition}, this submartingale process can be split into a predictable process $\Lambda_{srq}(t)$, and a residual martingale process $M_{srq}(t)$, that is
\begin{equation*}
N_{srq}(t)=\Lambda_{srq}(t)+M_{srq}(t).
\end{equation*}
If the derivative exists, the instantaneous hazard rate is defined by
\begin{equation*}
\lambda_{srq}(t) = \frac{d \Lambda_{srq}(t)}{dt},
\end{equation*}
and this is the object of modelling.

In particular, we define the multiplex relational event model as
\begin{equation}
\lambda_{srq}(t)= Y_{srq}(t) \, \lambda_{0q}(t) \, \exp\left(f_q(\boldsymbol{x}_{srq}(t)) +  \boldsymbol{\gamma}_q^\top \boldsymbol{z}_{srq}\right), \quad q \in Q.
\label{eq:mrem_hazard}
\end{equation}
Here the binary indicator $Y_{srq}(t)$ indicates if the event $(s, r, q)$ is at risk of occurring at time $t$. The baseline intensity function, $\lambda_{0q}(t)$, is common to all events of type $q$ and does not depend on the specific pair $(s, r)$. Covariate effects are included through $f_q(\boldsymbol{x}_{srq}(t))$, where covariates $\boldsymbol{x}_{srq}(t)$ may include both endogenous and exogenous covariates, measurable with respect to the history of the process. Endogenous covariates depend on the past history of the multiplex relational event process, whereas exogenous covariates are potentially time-varying actor-level attributes that are independent of previous interactions, such as gender or age. Covariates can enter the model linearly, i.e., $f_q(\boldsymbol{x}_{srq}(t)) = \boldsymbol{\beta}_q^\top \boldsymbol{x}_{srq}(t)$ for some parameter vector $\boldsymbol{\beta}_q$ in the case when all effects are linear, or non-linearly through a flexible smooth function $f_q(\cdot)$. In addition to these fixed effects, $\boldsymbol{z}_{srq}$ is a vector of covariates that indicates the inclusion of random effects in the model. In particular, this variable extracts  the random effects $\boldsymbol{\gamma}_{srq}$  for the interacting pair $(s,r)$ of type $q$ from the global vector of random effects $\boldsymbol{\gamma}_q$. As in standard mixed effect models, we assume that $\boldsymbol{\gamma}_{srq} \sim \mathcal{N}(\boldsymbol{0}, \boldsymbol{\Sigma}_q)$.

Equation~\eqref{eq:mrem_hazard} describes the intensity of the relational event process at each layer of the multiplex structure. Clearly, we expect some dependences across layers. The next section defines a number of interesting covariates to quantify these between-type effects. When the same covariate is present in multiple layers, irrespective of whether the covariate is a within or between-type one, the effects may also be shared across layers. For example, a covariate that enters the model with a linear effect may be associated to the same $\beta$ parameter across layers. Later on, we will refer to Equation~\eqref{eq:mrem_hazard} as the \textit{separable} formulation of the multiplex relational event model. This is the case when the component of the model related to the $q$ layer is associated with its own set of parameters $\boldsymbol{\theta}_q = (\lambda_{0q}, f_q, \boldsymbol{\Sigma}_{q})$. In contrast to this,  we will speak of the \textit{non-separable} case when there is some sharing across parameters. In the most extreme case, the models for each layer have the same set of covariates and the parameters associated to these  are all shared across layers. Then, the multiplex relational event model becomes
\begin{equation*}
\lambda_{srq}(t)
=
Y_{srq}(t)\,
\lambda_{0}(t)\,
\exp\left(
f(\boldsymbol{x}_{srq}(t))
+
\boldsymbol{\gamma}^\top \boldsymbol{z}_{srq}
\right), \quad \boldsymbol{\gamma}_{srq} \sim \mathcal{N}(\boldsymbol{0}, \boldsymbol{\Sigma}), \quad q \in Q.
\label{nonsep_hazard_shared}
\end{equation*}
This model is clearly more parsimonious than the separable one, though it may be too restrictive in practice.

\subsection{Multiplex network statistics} \label{sec:multiplex_statistics}
The multiplex relational event model is defined by a number of covariates, which describe the dynamics of the underlying process. As mentioned in the previous section, the covariates can be  either exogenous or endogenous. In this section, we focus in particular on the endogenous covariates, i.e., those that depend on the history of the multiplex process.
Table~\ref{tab:covariates} lists some popular ones from the relational event modelling literature \citep{boschi_introREM_2026}. 
These are split into those that describe nodal, dyadic and triadic effects, respectively. The third column of the table provides a formal definition of the covariates. This is given in terms of counts of events from the history of the process. Similar definitions can be given that give more weight to more recent observations or based on inter-arrival times \citep{juozaitiene_its_2024}. We will see examples of these in the simulation study in section~\ref{sec:simulation}.

\begin{table}[t]
    \centering 
    \makebox[\linewidth][c]{%
  \resizebox{1\textwidth}{!}{ 
    \begin{tabular}{p{6cm} p{4cm} p{12cm}}       
        \toprule
        ~~\\
        \textbf{EFFECT} & \textbf{VISUALIZATION} & \textbf{FORMAL DEFINITION} \\       
         ~~\\
        \toprule
               ~~\\
        \multicolumn{3}{l}{\textbf{NODAL}}\\
               ~~\\
  \raisebox{-1em}{\textbf{Sender Out-Degree}} & 
        \raisebox{-2em}{\begin{tikzpicture}[scale=0.4]        
        \draw[thick, draw=black, fill=blue!80, opacity=0.4] (1.5, 1.5) circle (0.4);
        
        \draw[thick, draw=black, fill=blue!80, opacity=0.4] (4.5, 2.5) circle (0.4);
        \draw[thick, draw=black, fill=blue!80, opacity=0.4] (4.5, 1.5) circle (0.4);
        \draw[thick, draw=black, fill=blue!80, opacity=0.4] (4.5, 0.5) circle (0.4);
        
        \draw[thick, ->, >=latex] (1.8, 1.8) -- (4.1, 2.5);
        \draw[thick, ->, >=latex] (1.9, 1.5) -- (4.1, 1.5);
        \draw[thick, ->, >=latex] (1.8, 1.2) -- (4.1, 0.5);
    
        \node at (1.5, 2.35) {$s$};
        \end{tikzpicture}}
        & \raisebox{-1em}{
        $      x^{(\mathrm{sd\text{-}out}, q^*)}_{sq}(t)= \displaystyle\sum_{\substack{t_i < t}} 1_{\{ s_i = s,\, q_i = q^*\}}$}\\
        \vspace{-0.8cm} \textbf{Sender In-Degree} & 
        \begin{tikzpicture}[scale=0.4]    
    \draw[thick, draw=black, fill=blue!80, opacity=0.4] (4.5, 1.5) circle (0.4);
    
    \draw[thick, draw=black, fill=blue!80, opacity=0.4] (1.5, 2.5) circle (0.4);
    \draw[thick, draw=black, fill=blue!80, opacity=0.4] (1.5, 1.5) circle (0.4);
    \draw[thick, draw=black, fill=blue!80, opacity=0.4] (1.5, 0.5) circle (0.4);
    
    \draw[thick, ->, >=latex] (1.9, 2.5) -- (4.2, 1.8);
    \draw[thick, ->, >=latex] (1.9, 1.5) -- (4.2, 1.5);
    \draw[thick, ->, >=latex] (1.9, 0.5) -- (4.2, 1.2);

    \node at (4.5, 2.3) {$s$};
\end{tikzpicture}
         & \vspace{-0.8cm} $x^{(\mathrm{sd\text{-}in}, q^*)}_{sq}(t) \, \, \, =\displaystyle\sum_{\substack{t_i < t}}  1_{\{r_i = s, \, q_i=q^*\}}$ \\
        \raisebox{-1em}{\textbf{Receiver Out-Degree}} & 
        \raisebox{-2em}{\begin{tikzpicture}[scale=0.4]
        \draw[thick, draw=black, fill=blue!80, opacity=0.4] (1.5, 1.5) circle (0.4);
        
        \draw[thick, draw=black, fill=blue!80, opacity=0.4] (4.5, 2.5) circle (0.4);
        \draw[thick, draw=black, fill=blue!80, opacity=0.4] (4.5, 1.5) circle (0.4);
        \draw[thick, draw=black, fill=blue!80, opacity=0.4] (4.5, 0.5) circle (0.4);
        
        \draw[thick, ->, >=latex] (1.8, 1.8) -- (4.1, 2.5);
        \draw[thick, ->, >=latex] (1.9, 1.5) -- (4.1, 1.5);
        \draw[thick, ->, >=latex] (1.8, 1.2) -- (4.1, 0.5);
    
        \node at (1.5, 2.35) {$r$};
        \end{tikzpicture}}
        & \raisebox{-1em}{$x^{(\mathrm{rd\text{-}out}, q^*)}_{rq}(t)= \displaystyle\sum_{\substack{t_i < t}} 1_{\{ s_i = r,\, q_i = q^*\}}$}\\
        \vspace{-0.8cm} \textbf{Receiver In-Degree} & 
        \begin{tikzpicture}[scale=0.4]
    \draw[thick, draw=black, fill=blue!80, opacity=0.4] (4.5, 1.5) circle (0.4);
    
    \draw[thick, draw=black, fill=blue!80, opacity=0.4] (1.5, 2.5) circle (0.4);
    \draw[thick, draw=black, fill=blue!80, opacity=0.4] (1.5, 1.5) circle (0.4);
    \draw[thick, draw=black, fill=blue!80, opacity=0.4] (1.5, 0.5) circle (0.4);
    \draw[thick, ->, >=latex] (1.9, 2.5) -- (4.2, 1.8);
    \draw[thick, ->, >=latex] (1.9, 1.5) -- (4.2, 1.5);
    \draw[thick, ->, >=latex] (1.9, 0.5) -- (4.2, 1.2);
    \node at (4.5, 2.3) {$r$};
\end{tikzpicture}
         & \vspace{-0.8cm} $x^{(\mathrm{rd\text{-}in}, q^*)}_{rq}(t) \, \, \, =\displaystyle\sum_{\substack{t_i < t}}  1_{\{r_i = r, \, q_i=q^*\}}$ \\
         \hline
         ~~\\
         \multicolumn{3}{l}{\textbf{DYADIC}}\\
        \vspace{-0.3cm} \textbf{Repetition} & 
        \begin{tikzpicture}[scale=0.4]
        \draw[thick, draw=black, fill=blue!80, opacity=0.4] (1.4, 1.5) circle (0.4);
        
        \draw[thick, draw=black, fill=blue!80, opacity=0.4] (4.5, 1.5) circle (0.4);

        \draw[thick, ->, >=latex] (1.8, 1.3) to[out=0, in=180] (4.2, 1.3);
        \draw[thick, ->, dashed, >=latex] (1.8,1.6) to[out=0, in=180] (4.2, 1.6);

        \node at (1.5, 2.3) {$s$};
        \node at (4.5, 2.3) {$r$};
        \end{tikzpicture}
        & \vspace{-0.5cm} $x_{srq}^{(\mathrm{rep}, q^*)}(t) = \displaystyle\sum_{\substack{t_i < t}} 1_{\{s_i = s, \, r_i = r, \, q_i=q^*\}}$\\    
        \vspace{-0.5cm} \textbf{Retaliation/Reciprocity} & 
        \begin{tikzpicture}[scale=0.4]
        \draw[thick, draw=black, fill=blue!80, opacity=0.4] (1.5, 1.5) circle (0.4);
        
        \draw[thick, draw=black, fill=blue!80, opacity=0.4] (4.6, 1.6) circle (0.4);     

        \draw[thick, ->, dashed, >=latex] (1.9,1.6) to[out=0, in=180] (4.2, 1.6);
        \draw[thick, ->, >=latex]  (4.3, 1.3)  to[out=180, in=0] (1.8, 1.3);

        \node at (1.5, 2.3) {$s$};
        \node at (4.5, 2.3) {$r$};
        \end{tikzpicture}
        & \vspace{-0.5cm} $x^{(\mathrm{rec}, q^*)}_{srq}(t)=\displaystyle\sum_{t_i < t} 1_{\{s_i = r, \, r_i = s, \, q_i = q^*\}}$   \\
        \hline
        \\
        \multicolumn{3}{l}{\textbf{TRIADIC}}\\  
        \vspace{-0.8cm}
        \textbf{Transitive Closure} &
        \begin{tikzpicture}[scale=0.4]
        \draw[thick, draw=black, fill=blue!80, opacity=0.4] (0, 0) circle (0.4);
        \draw[thick, draw=black, fill=blue!80, opacity=0.4] (3, 0) circle (0.4);    
        \draw[thick, draw=black, fill=blue!80, opacity=0.4] (1.5, 2.3) circle (0.4);
    
        \node at (-0.2, 0.8) {$s$};
        \node at (3.3, 0.8) {$r$};
        \node at (1.5, 3.2) {$a$};
    
        \draw[thick, ->, >=latex] (1.8, 2) -- (2.8, 0.3); 
        \draw[thick, ->, >=latex] (0.2, 0.4) -- (1.2, 2.1); 
    
        \draw[thick, dashed, ->, >=latex] (0.5, 0) to[out=0, in=180] (2.6, 0); 
         \end{tikzpicture}
        & \vspace{-1.2cm} $x^{(\mathrm{tc}, q^*)}_{srq}(t)  = \displaystyle \sum_{\substack{a \neq s, r}} \min \Biggl\{ \sum_{t_i < t} 1_{\{ s_i = s, \, r_i = a, \, q_i=q^* \}}, \, \sum_{t_i < t} 1_{\{s_i = a, \, r_i = r, q_i=q^* \}} \Biggl\}$  \\
        \vspace{-1cm}
        \textbf{Cyclic Closure} &
        \begin{tikzpicture}[scale=0.4]
        \draw[thick, draw=black, fill=blue!80, opacity=0.4] (0, 0) circle (0.4);
        \draw[thick, draw=black, fill=blue!80, opacity=0.4] (3, 0) circle (0.4);    
        \draw[thick, draw=black, fill=blue!80, opacity=0.4] (1.5, 2.3) circle (0.4);
    
        \node at (-0.2, 0.8) {$s$};
        \node at (3.3, 0.8) {$r$};
        \node at (1.5, 3.2) {$a$};
    
        \draw[thick, ->, >=latex] (2.8, 0.3) -- (1.8, 2.1); 
        \draw[thick, ->, >=latex] (1.2, 2) -- (0.2, 0.4); 
    
        \draw[thick, dashed, ->, >=latex] (0.5, 0) to[out=0, in=180] (2.6, 0); 
         \end{tikzpicture}
         & \vspace{-1.2cm} $x^{(\mathrm{cc}, q^*)}_{srq}(t)  = \displaystyle \sum_{\substack{a \neq s, r}} \min \Biggl\{ \sum_{t_i < t} 1_{\{ s_i = r, \, r_i = a, \, q_i=q^* \}}, \, \sum_{t_i < t} 1_{\{ s_i = a, \,  r_i = s, \, q_i=q^* \}} \Biggl\}$     \\
        \vspace{-1cm}
        \textbf{Sender Balance} &
        \begin{tikzpicture}[scale=0.4]
        \draw[thick, draw=black, fill=blue!80, opacity=0.4] (0, 0) circle (0.4);
        \draw[thick, draw=black, fill=blue!80, opacity=0.4] (3, 0) circle (0.4);    
        \draw[thick, draw=black, fill=blue!80, opacity=0.4] (1.5, 2.3) circle (0.4);
    
        \node at (-0.2, 0.8) {$s$};
        \node at (3.3, 0.8) {$r$};
        \node at (1.5, 3.2) {$a$};
    
        \draw[thick, ->, >=latex] (1.8, 2) -- (2.8, 0.3); 
        \draw[thick, ->, >=latex] (1.2, 2) -- (0.2, 0.3); 
    
        \draw[thick, dashed, ->, >=latex] (0.5, 0) to[out=0, in=180] (2.6, 0); 
         \end{tikzpicture}
        & \vspace{-1.2cm} $x^{(\mathrm{sb}, q^*)}_{srq}(t)  = \displaystyle \sum_{\substack{a \neq s, r}} \min \Biggl\{ \sum_{t_i<t} 1_{\{ s_i = a, \, r_i = s, \, q_i=q^* \}}, \, \sum_{t_i<t} 1_{\{ s_i = a, \, r_i = r,\, q_i=q^* \}} \Biggl\}$   \\
        \vspace{-1cm}
        \textbf{Receiver Balance} & 
        \begin{tikzpicture}[scale=0.4]
        \draw[thick, draw=black, fill=blue!80, opacity=0.4] (0, 0) circle (0.4);
        \draw[thick, draw=black, fill=blue!80, opacity=0.4] (3, 0) circle (0.4);    
        \draw[thick, draw=black, fill=blue!80, opacity=0.4] (1.5, 2.3) circle (0.4);
    
        \node at (-0.2, 0.8) {$s$};
        \node at (3.3, 0.8) {$r$};
        \node at (1.5, 3.2) {$a$};
    
        \draw[thick, ->, >=latex] (2.8, 0.3) --  (1.8, 2); 
        \draw[thick, ->, >=latex] (0.2, 0.4) -- (1.2, 2); 
    
        \draw[thick, dashed, ->, >=latex] (0.5, 0) to[out=0, in=180] (2.6, 0); 
         \end{tikzpicture}
         & \vspace{-1.2cm} $x^{(\mathrm{rb}, q^*)}_{srq}(t)  = \displaystyle \sum_{\substack{a \neq s, r}} \min \Biggl\{ \sum_{t_i < t} 1_{\{ s_i = s, \, r_i = a, \, q_i = q^* \}}, \sum_{t_i< t} 1_{\{ s_i = r, \, r_i = a, \, q_i=q^* \}} \Biggl\}$    \\
        \bottomrule
    \end{tabular}}}
    \caption{Endogenous covariates describing effects on the intensity of the multiplex REM at layer $q$ based on the history of $q^*$. Solid lines ($\to$) refer to past relational events, while dashed arrows ($\dashrightarrow$) indicate current relational events. When  $q \ne q^*$, the covariates capture cross-layer effects.}
    \label{tab:covariates}
\end{table}

The definitions are given for an interaction $(s,r)$  in layer $q$ at a generic time $t$ with respect to the history of a process of type $q^*$.  Therefore, while the case $q=q^*$ refers to a \textit{within-type} endogenous covariate, such as those that are included in standard relational event models, the case $q\ne q^*$ corresponds to a covariate that captures cross-layer effects. Via one of these \textit{between-type} endogenous covariates, the history of the $q^*$ process influences the instantaneous rate of an interaction of type $q$, thereby capturing cross-layer dependences among the processes.
Referring back to the gossiping example of section~\ref{sec:motivating_example}, a within-type repetition effect in the self-reported gossip model, say, would measure whether past instances of actor $s$ gossiping about actor $r$ has an effect on the rate that $s$ gossips about $r$ again at a particular point in time. 
On the other hand, a between-type reciprocity effect from perceived to self-reported gossip would measure whether past instances of actor $s$ perceiving that $r$ is gossiping about $s$ has an effect on $s$ reporting gossiping about $r$ in the future, thereby capturing a dependence between the two processes. 

\section{Inference} \label{sec:inference}

In this section, we extend existing inferential techniques for relational event models to the case of multiplex relational event models. We consider three observational settings: the case of fully observed sequences of events across all layers (section~\ref{sec:full_data}), the case of event data aggregated across time intervals (section~\ref{sec:int_data}), and, finally, the case of censored counts, whereby for each time interval we only know whether at least one event of each type occurred (section~\ref{sec:interval_rightcensoring}). Each setting leads to a different likelihood formulation, although we will see how inference can be performed via flexible generalized additive models in all cases.

\subsection{Complete data}\label{sec:full_data}

We assume that the full sequence of time ordered events $(s_i,r_i,q_i,t_i)$, $i=1,\ldots, n$, is observed. Then, the likelihood of the multiplex relational event data is given by
\begin{equation}
L(\boldsymbol{\theta})
=
\prod_{i=1}^{n}
\bigg(
\sum_{(s,r,q)\in R(t_i)} \lambda_{srq}(t_i)
\bigg)
\exp\bigg\{
-
\sum_{(s,r,q)\in R(t_i)}
\int_{t_{i-1}}^{t_i}
\lambda_{srq}(\tau)\, d\tau
\bigg\}
\,
\frac{\lambda_{s_i r_i\, q_i}(t_i)}{\sum_{(s,r,q)\in R(t_i)} \lambda_{sr q}(t_i)}
\label{eq:full_like},
\end{equation}
where $R(t_i)$ denotes the set of  relational events across all event types that are at risk of occurring at time $t_i$, and $\lambda_{srq}(t)$ is the corresponding intensity from the multiplex relational event model~\eqref{eq:mrem_hazard}. For each event type $q\in Q$, the intensity is parametrized by $\boldsymbol{\theta}_q=(\lambda_{0q}, f_q, \boldsymbol{\Sigma}_q)$, and $\boldsymbol{\theta} = \{\boldsymbol{\theta}_q\}_{q \in Q}$ denotes the full set of parameters. In the case of a non-separable model, we slightly abuse the notation, as, in this case, the vector of parameters $\boldsymbol{\theta}$  includes shared parameters only once across the layers in which they are shared. 

As in the case of standard relational event models, maximizing the full likelihood~\eqref{eq:full_like} is computationally challenging both due to the integrals of complex time-varying hazard functions and to the sums over a potentially large risk set. A common approach from the survival modelling literature \citep{cox_1975} is to consider only the part of the likelihood which conditions on an event occurring at time $t_i$. For the multiplex REM likelihood, this yields the partial likelihood:
\begin{equation}
L_P(\boldsymbol{\theta}) = 
\prod_{i=1}^{n} 
\frac{
\lambda_{s_i r_i q_i}(t_i)
}{
\sum_{(s,r, q) \in R(t_i)} \lambda_{s r q}(t_i)
}
= 
\prod_{i=1}^{n} 
\frac{
\lambda_{0q_i}(t_i) \exp \left(f_{q_i}(\boldsymbol{x}_{s_ir_iq_i}(t_i)) +  \boldsymbol{\gamma}^\top \boldsymbol{z}_{s_ir_iq_i}\right) 
}{
\sum_{(s,r,q) \in R(t_i)} \lambda_{0q}(t_i) \exp
\left(f_q(\boldsymbol{x}_{srq}(t_i)) +  \boldsymbol{\gamma}^\top \boldsymbol{z}_{srq}\right)
}
\label{eq:pl}
\end{equation}
This formulation simplifies inference as it does not require evaluation of the integrals. However, calculating the denominator remains computationally demanding in realistic networks where the risk set $R(t_i)$ can be very large. 

Similarly to standard relational event models \citep{VU2015121}, we address this limitation with the use of nested case–control sampling \citep{Borgan1995Methods}. In particular, for each observed event $(s_i, r_i, q_i, t_i)$, a non-event $(s_i^*, r_i^*, q_i^*, t_i)$ is randomly sampled from the risk set $R(t_i)$, with $s_i \ne s_i^*$ and $r_i \ne r_i^*$. Let 
\[\eta_i = f_{q_i}(\boldsymbol{x}_{s_ir_iq_i}(t_i)) + \boldsymbol{\gamma}^\top \boldsymbol{z}_{s_ir_iq_i}\]
and 
\[\eta_i^* = f_{q_i^*}(\boldsymbol{x}_{s_i^*r_i^*q_i^*}(t_i)) + \boldsymbol{\gamma}^\top \boldsymbol{z}_{s_i^*r_i^*q_i^*}\]
denote the linear predictors for the event and non-event, respectively. We consider the sampled partial likelihood:
\begin{align}
\tilde{L}_P(\boldsymbol{\theta}) &=
\prod_{i=1}^{n}
\frac{
\lambda_{0q_i}(t_i) \exp(\eta_i)
}{
\lambda_{0q_i}(t_i) \exp(\eta_i) + \lambda_{0q_i^*}(t_i) \exp(\eta_i^*)
}
\nonumber \\
&= \prod_{i=1}^{n}
\frac{
\exp\!\left(\ln \lambda_{0q_i}(t_i) - \ln \lambda_{0q_i^*}(t_i) + \eta_i - \eta_i^*\right)
}{ 
1 + \exp\!\left(\ln \lambda_{0q_i}(t_i) - \ln \lambda_{0q_i^*}(t_i) + \eta_i - \eta_i^*\right)
}, \nonumber
\end{align}
which further conditions on only the event and non-event potentially occurring at time $t_i$.

The fact that each observation enters the likelihood via the familiar sigmoid function means that this partial likelihood is equivalent to that of a degenerate logistic regression, i.e., a logistic regression where the responses are all set to one. This is also the case for standard relational event models and it has allowed the fitting of complex generalized additive models to relational event networks \citep{boschi_smooth_2025,filippi-mazzola2024}. For a multiplex relational event model, the construction of the model specification needs to account for the multiplex nature of the data. Considering for simplicity the case of linear effects, an observation that has been matched to a non-event from the same layer will contribute to the design matrix of the model with the vector $(\boldsymbol{x}_{s_ir_iq_i}(t_i)-\boldsymbol{x}_{s_i^*r_i^*q_i}(t_i),\textbf{0})$. If $q_i^* \ne q_i$ and the model is separable, then it will contribute with $(\boldsymbol{x}_{s_ir_iq_i}(t_i),-\boldsymbol{x}_{s_i^*r_i^*q_i^*}(t_i))$. Between these two extremes, one can have the case where one or more parameter are shared across layers. This can be handled in a similar way to the $q_i^* = q_i$ case, although the $q_i^*$ covariate of the non-event is now calculated according to that process. Note how the logistic regression model does not include an intercept, since intercepts are not included in the parametric part of the multiplex relational event model~\eqref{eq:mrem_hazard}. Regarding the baseline hazards, these  cancel out from the partial likelihood when $q_i^* = q_i$, as they are evaluated at the same time point $t_i$. This is not the case when  $q_i^* \ne q_i$, which would make the difference of the baseline log-intensities estimable. In order to estimate the baseline across all layers, time-shifting can be conducted prior to the fitting of the multiplex relational event model, as discussed by \cite{lembo_relational_2025} for the case of relational event models.


The approach described above corresponds to the fitting of a single logistic regression additive model. This is generally efficient, and certainly much more efficient than standard survival modelling approaches based on the full risk set. However, the design matrix can be quite large: the number of rows is given by the full sequence of events and the number of columns by the sum of the covariates across all layers. Both can be potentially large. When the multiplex model is separable, inference simplifies significantly if we consider a more restrictive part of the likelihood where the non-event is sampled from the same layer. That is,  we consider the partial likelihood
\begin{align}
\tilde{L}_P^{\mathrm{sep}}(\boldsymbol{\theta}) &= \prod_{i=1}^{n}
\frac{
\exp\!\left( f_{q_i} \big(\boldsymbol{x}_{s_ir_iq_i}(t_i)\big) - f_{q_i} \big(\boldsymbol{x}_{s_i^*r_i^*q_i}(t_i)\big) + \boldsymbol{\gamma}^\top \big(\boldsymbol{z}_{s_i r_i q_i} -  \boldsymbol{z}_{s_i^* r_i^* q_i}) \right)
}{ 
1 + \exp\!\left(f_{q_i} \big(\boldsymbol{x}_{s_ir_iq_i}(t_i)\big) - f_{q_i} \big(\boldsymbol{x}_{s_i^*r_i^*q_i}(t_i)\big) + \boldsymbol{\gamma}^\top \big(\boldsymbol{z}_{s_i r_i q_i} -  \boldsymbol{z}_{s_i^* r_i^* q_i})\right)
}, \nonumber
\end{align}
with $q_i=q_i^*$ for all $i$. This can be re-written as a product of likelihoods across the different layers
\begin{align}
\tilde{L}_P^{\mathrm{sep}}(\boldsymbol{\theta}) &= \prod_{q \in Q}\prod_{\substack{i=1 \\ q_i = q}}^n
\frac{
\exp\!\left( f_{q_i} \big(\boldsymbol{x}_{s_ir_iq_i}(t_i)\big) - f_{q_i} \big(\boldsymbol{x}_{s_i^*r_i^*q_i}(t_i)\big) + \boldsymbol{\gamma}^\top \big(\boldsymbol{z}_{s_i r_i q_i} -  \boldsymbol{z}_{s_i^* r_i^* q_i}) \right)
}{ 
1 + \exp\!\left(f_{q_i} \big(\boldsymbol{x}_{s_ir_iq_i}(t_i)\big) - f_{q_i} \big(\boldsymbol{x}_{s_i^*r_i^*q_i}(t_i)\big) + \boldsymbol{\gamma}^\top \big(\boldsymbol{z}_{s_i r_i q_i} -  \boldsymbol{z}_{s_i^* r_i^* q_i})\right)
}. \nonumber
\end{align}
Since each layer is modelled by its own parameters $\boldsymbol{\theta}_q$, this partial likelihood can be maximized by $|Q|$ separate logistic additive models. Importantly, these can be fitted independently for each event type and the design matrix of each of these models is now significantly smaller in size.


\subsection{Data aggregated across time intervals}\label{sec:int_data}

In many applications, measurements of the individual events are not available, but only the counts of the events within prespecified time intervals, such as days or weeks. Surveys are an example of this, where respondents may be asked how many times a specific interaction occurred within a certain period of time. For simplicity, we assume that this aggregation applies to all layers of the multiplex process. Thus, the likelihood of the multiplex relational event process will now be in terms of the number of events of each type occurring within each observation interval.


In order to formalize this, let $K$ denote the number of observational intervals, and $(t_{k-1}, t_k]$, for $k=1,\ldots,K$, the actual intervals. Let $N_{srqk}$ denote the number of events associated with sender $s$, receiver $r$, and event type $q$ in interval $k$. 
Conditional on the history $\mathcal{H}_{t_{k-1}}$ of the multiplex relational event process at time $t_{k-1}$, we have
\begin{equation}
N_{srqk} = N_{srq}(t_k) - N_{srq}(t_{k-1}) \;\big|\; \mathcal{H}_{t_{k-1}}
\sim
\text{Poisson}\!\left(\int_{t_{k-1}}^{t_k} \lambda_{srq}{(t)} dt\right),
\label{eq:int_poisson}
\end{equation}
where $\lambda_{srq}{(t)}$ is linked to the covariates via the multiplex relational event model in equation~\eqref{eq:mrem_hazard}.
Since the integral is generally intractable and the covariates are usually observed only at the extremes of a time interval, we approximate this integral by setting the intensity to a constant value within the interval. As discussed in \cite{poda2025hypereventnetworkmodellingpartially}, a good choice is to set each covariate  to the average between its value at the beginning and at the end of the interval. In some cases, the value at the beginning of the interval may be more appropriate, as we will see in the school survey data analysis in section~\ref{sec:realdata}. Assuming, without loss of generality, that all variables are set to the average within each interval, the integral of equation~\eqref{eq:int_poisson} is approximated by
\begin{equation}
\lambda_{srqk}(\boldsymbol{\theta}_q)=(t_k - t_{k-1})\lambda_{0q}\big(\frac{{t}_{k-1}+t_k}{2}\big)   \,
\exp\left(f_q\big(\frac{\boldsymbol{x}_{srq}(t_{k-1})+\boldsymbol{x}_{srq}(t_{k})}{2}\big) +  \boldsymbol{\gamma}^\top \boldsymbol{z}_{srq}\right).
\label{eq:int_approx}
\end{equation}

Given observed counts $n_{srqk}$, across all intervals $k=1, \ldots, K$ and layers $q \in Q$, the full likelihood is then given by
\begin{align}
L(\boldsymbol{\theta}) &= \prod_{k=1}^K 
\, \prod_{(s,r,q) \in R(k)} 
e^{-\lambda_{srqk}(\boldsymbol{\theta}_q)}
\frac{(\lambda_{srqk}(\boldsymbol{\theta}_q))^{n_{srqk}}} {n_{srqk}!}, \label{eq:poisson_full}
\end{align}
where $R(k)$ denotes the set of events that are at risk of occurring in interval $k$. This can simply be fitted by a Poisson regression additive model with offsets equal to $\log(t_k-t_{k-1})$. If the multiplex relational event model is non-separable, then the design matrix will only have one column for any covariate whose parameter is shared across layers. This column will contain measurements of that covariate across the different layers and for all elements of the risk set. For the covariates whose parameters are not shared, the design matrix will instead contain separate columns across the different layers, where the elements of each vector are set to zero outside of the corresponding layer. If all parameters are event-specific, e.g., when the model is separable,  maximization can be conducted efficiently via  $|Q|$ independent Poisson regression additive models. This is because the likelihood~\ref{eq:poisson_full} can be rewritten as 
\begin{align}
L(\boldsymbol{\theta}) &= \prod_{q\in Q}   \prod_{k=1}^K 
\, \prod_{(s,r) \in R_q(k)} 
e^{-\lambda_{srqk}(\boldsymbol{\theta}_q)}
\frac{(\lambda_{srqk}(\boldsymbol{\theta}_q))^{n_{srqk}}} {n_{srqk}!},\nonumber
\end{align}
with $R_q(k)$ denoting the risk set specific to event type $q$. Thus, it factorizes directly across the event types. 

\subsection{Censored counts across time intervals}
\label{sec:interval_rightcensoring}

We now consider the setting where we know only whether at least one event happened in a given time interval. This tends to occur less frequently than  aggregated counts, but is not uncommon in sociological studies. In fact, it is precisely the case of the survey data in section~\ref{sec:motivating_example}, where  we only have information about whether at least one gossiping event occurred within the time interval covered by each wave. As in section~\ref{sec:int_data}, we therefore have data aggregated per time interval --- here, the four waves of the study --- but the event counts are \textit{right-censored at 1}. 
Formally, what we observe is 
\begin{equation*}
W_{srqk} = 1_{\{N_{srqk} > 0\}}.
\end{equation*}
Thus, given the underlying multiplex relational event process, the likelihood is given by
\begin{equation}
L(\boldsymbol{\theta}) = \prod_{k=1}^K \prod_{(s,r,q) \in R(k)} 
\underbrace{\left( e^{-\int_{t_{k-1}}^{t_k} \lambda_{srq}(u) \, du} \right)^{1 - w_{srqk}}}_{\raisebox{-0.7em}{\footnotesize $n_{srqk} \, = \, 0$}} 
\underbrace{\left(1 - e^{-\int_{t_{k-1}}^{t_k} \lambda_{srq}(u) \, du}\right)^{w_{srqk}}}_{\raisebox{-0.7em}{\footnotesize $n_{srqk} \, > \, 0$}}.
\label{eq:fulllik_int_cens}
\end{equation}
This, using the same approximation~\eqref{eq:int_approx} of the integrals, becomes
\begin{equation}
L(\boldsymbol{\theta}) \approx \prod_{k=1}^K \prod_{(s,r,q) \in R(k)} 
\left( e^{-\lambda_{srqk}(\boldsymbol{\theta}_q)} \right)^{1 - w_{srqk}} 
\left(1 - e^{-\lambda_{srqk}(\boldsymbol{\theta}_q)}\right)^{w_{srqk}},
\label{eq:int_cens}
\end{equation}
which corresponds to a right-censored Poisson regression additive model with offsets $\log(t_k - t_{k-1})$. Since the right-censoring is at $1$, this is equivalent to a Binomial regression additive model with complementary log-log (cloglog) link. This is discussed in \cite{poda2025hypereventnetworkmodellingpartially} for the case of a standard relational hyperevent model. Moreover, as in section~\ref{sec:int_data}, the likelihood~\eqref{eq:int_cens} can be written as products across event types. Thus, as in that case, $|Q|$ independent Binomial or censored Poisson regression models can be fitted when the multiplex relational event model is separable. 
\subsection{Efficient implementation of multiplex REMs} \label{sec:implementation}

A key advantage of the inferential approaches derived in the previous three sections is that they can be efficiently implemented using standard statistical software. In particular, they only require an efficient implementation of generalized additive models, such as that in the \texttt{R} package \texttt{mgcv} \citep{wood2017}. Depending on the observational setting, estimation can be performed using logistic regression (complete data, section~\ref{sec:full_data}), Poisson regression (aggregated counts, section~\ref{sec:int_data}), censored Poisson or cloglog Binomial (censored counts, section~\ref{sec:interval_rightcensoring}). 
Flexible models can be contemplated with a mix of fixed, smooth and random effects, which allow to capture the dynamics of complex multiplex relational event processes. 

Model selection approaches are also readily available in order to compare models of different complexity and to highlight the most informative covariate processes. In the case of generalized additive models, one can use the double-penalty approach of \citet{penaltySelectTrue}, which is also implemented in the \texttt{mgcv} package. This includes an additional penalty on the smooth terms, so that they are shrunk entirely to zero if uninformative.  If the models have only linear effects, and a constant baseline in the case of aggregated or censored counts, more traditional information criteria, such as the Akaike Information Criterion (AIC) or the Bayesian Information Criterion (BIC), can be used. These are available in most packages for fitting generalized linear models. Given a sample of $n$ events across layers and a multiplex relational model with $k$ parameters in total, the two criteria are given by
\[
\text{AIC} =  - 2 \ln(L(\hat{\boldsymbol{\theta}}))+2 k, \qquad
\text{BIC} =  - 2 \ln(L(\hat{\boldsymbol{\theta}})) +\ln(n)k,
\]
respectively, with $\hat{\boldsymbol{\theta}}$ the maximum likelihood estimates of the parameters.
When the parameters are estimated independently for each event type $q \in Q$, the criteria can be calculated from the fitting of the individual models at each layer. In particular, if $k_q$ is the number of parameters in layer $q$, $\hat{\boldsymbol{\theta}}_q$ the estimated parameters and  $n_q$ the number of events, the two criteria are given by
\[
\text{AIC} = - 2 \sum_{q\in Q} \ln(L(\hat{\boldsymbol{\theta}}_q) + 2 \sum_{q\in Q} k_q, \qquad
\text{BIC} =  - 2 \sum_{q\in Q} \ln(L(\hat{\boldsymbol{\theta}}_q)) +\ln(n)\sum_{q\in Q} k_q,
\]
respectively.
Notice how the total AIC is additive across event types, so it can be simply calculated by adding the AIC of the models for each layer, while the BIC is not additive since $\ln(\sum_{q \in Q}n_q) \neq \sum_{q\in Q}\ln(n_q)$. The likelihood is, however, additive, as discussed earlier. So also the BIC can be calculated efficiently from the fitted models in each layer.

\section{Simulation study}\label{sec:simulation}

This section evaluates the proposed methodology via a simulation study. The first study evaluates the effect of ignoring between-type mechanisms in a multiplex relational event process by fitting standard relational event models without cross-layer effects (section~\ref{sec:sim_1}). The second study evaluates the effect of using a separable model when the underlying multiplex relational event process is non-separable (section~\ref{sec:sim_2}).

\subsection{Cross-network generating mechanism}\label{sec:sim_1}
We simulate relational events from a multiplex relational event process with two layers, driven exclusively by a between-type reciprocity mechanism which acts in both directions. Denoting with $q$ and $q^*$ the two event types, we set the corresponding intensity processes to
\begin{equation}
\begin{aligned}
\lambda_{srq}(t)
&=
Y_{srq}(t)\,
\lambda_{0q}
\exp\Big\{ \beta_q \,
x^{(rec,q^*)}_{srq}(t)
\Big\}, \\ 
\lambda_{srq^*}(t)
&=
Y_{srq^*}(t)\,
\lambda_{0q^*}
\exp\Big\{ \beta_{q^*} \,
x^{(rec,q)}_{srq^*}(t)
\Big\},
\label{eq:model_sim1}
\end{aligned}
\end{equation}
respectively. We set the baseline function  to the constants $\lambda_{0q}=0.15$ and $\lambda_{0q^*}=0.17$ for the two processes, respectively,  and the reciprocity effects to $\beta_q=0.9$ and  $\beta_q^*=0.7$, respectively.  Since both the baseline and the parameter associated to the endogenous covariate are different across the two layers, this multiplex relational event model is an example of a separable model. As for the between-type reciprocity covariate, we adapt the definition of Table~\ref{tab:covariates} to give more weight to the most recent observations. In particular, we compute this covariate as an exponentially decayed sum of past reciprocated events in the other network, i.e., 
\begin{equation*}
    x_{srq}^{(rec, q^*)}(t)= \sum_{t^*_{rsq^*} < t} e^{-\delta (t-t^*_{rsq^*})},
\end{equation*}
where \(t^*_{rsq^*}\) denotes the times of  reciprocated events of type \(q^*\)  prior to time $t$ and the rate of temporal memory decay $\delta$ is set to $1$.


A between-type reciprocity mechanism that acts in both directions may mimic the case of a within-type repetition effect. This could be captured by two separate relational event models at each layer: for the events in layer $q$ the model includes the within-type repetition covariate
\[x_{srq}^{(rep, q)}(t) = \sum_{t_{srq}^* < t} e^{-\frac{\delta}{2} (t - t_{srq}^*)},\]
while $x_{srq^*}^{(rep, q^*)}(t)$ is used for the events in layer $q^*$.
For these covariates, the decay rate is set to half of that of reciprocity, as it takes two reciprocal events to create a fake repetition effect. We expect that such a misspecified model  may not fit as well as the correctly specified multiplex relational event model, where the between-type reciprocity covariate $x_{srq}^{(rec, q^*)}$ is used in layer $q$ and $x_{srq^*}^{(rec, q)}$ is instead used in layer $q^*$. 

\begin{figure}[t]
    \centering
    \begin{subfigure}{0.48\textwidth}
        \centering      
        
        \includegraphics[width=\textwidth]{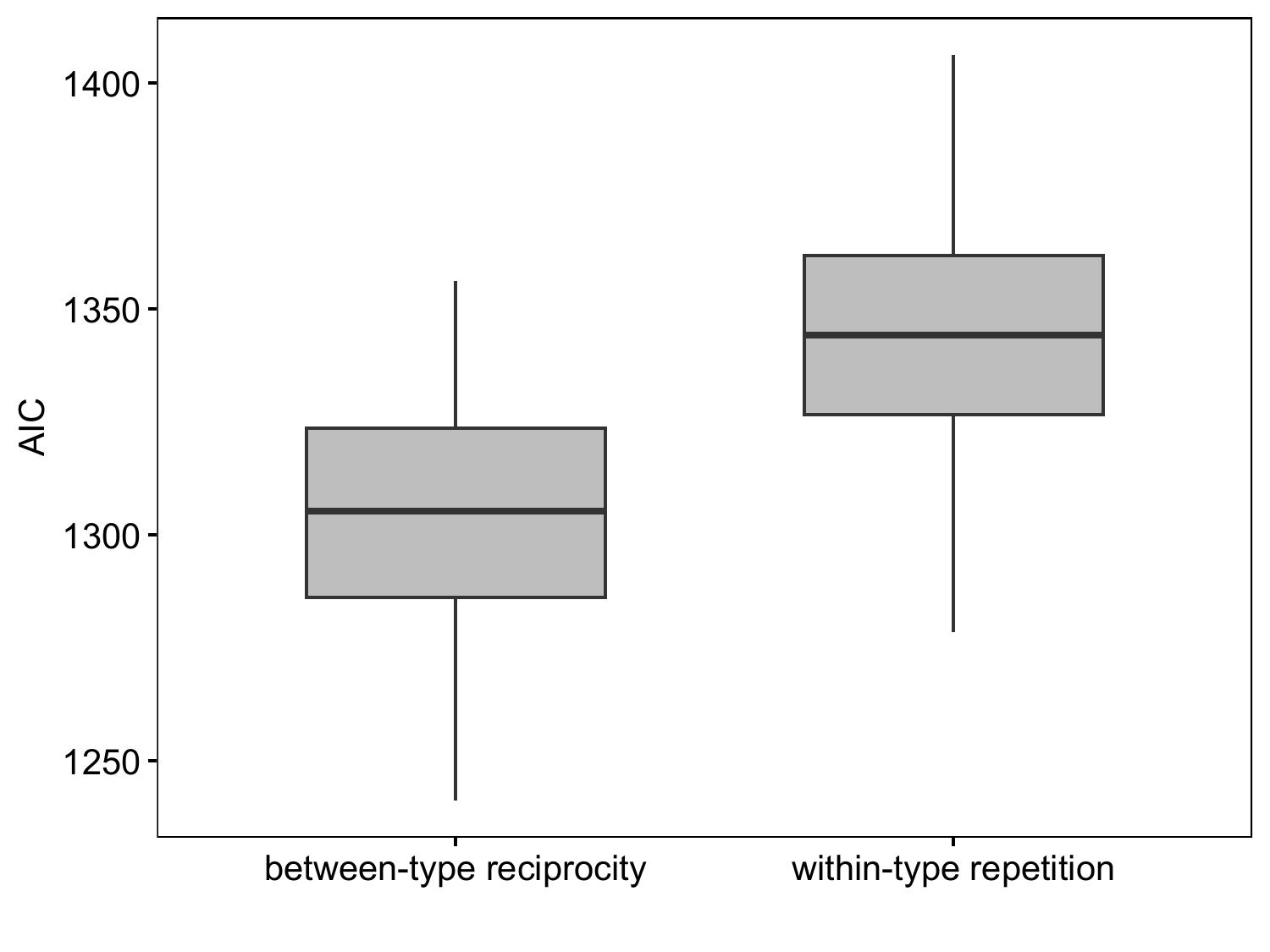}
        \caption{Cross-network generating mechanism}
        \label{fig:sim1_AIC}
    \end{subfigure}
    \hfill
    \begin{subfigure}{0.48\textwidth}
        \centering      
        \includegraphics[width=\textwidth]{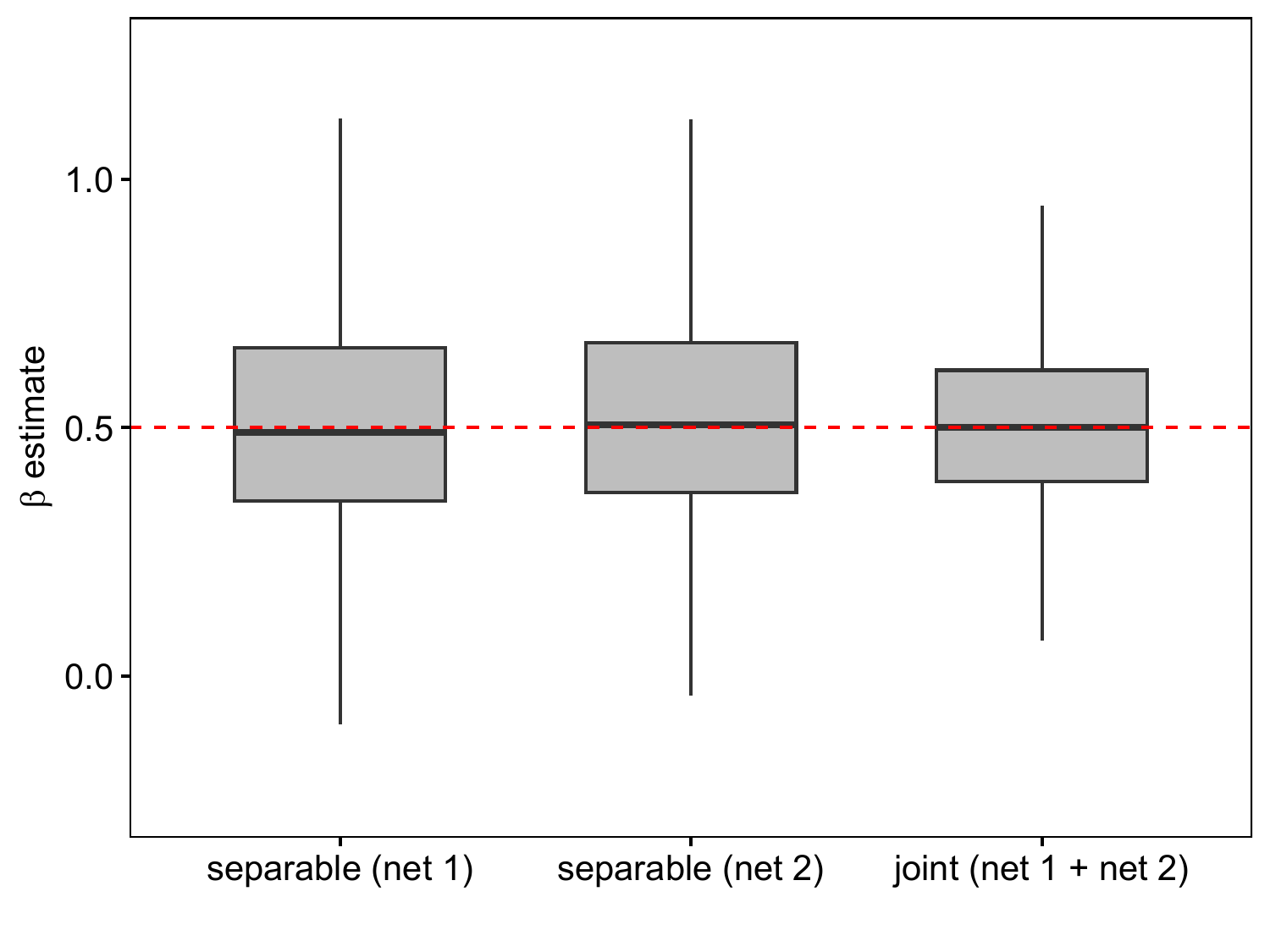}
        \caption{Non-separable generating mechanism}
        \label{fig:sim2_beta}
    \end{subfigure}
    
    \caption{(a) AIC values across $50$ simulations for a correctly specified multiplex REM with a between-type reciprocity effect and a misspecified model with a within-type repetition effect. Accounting for cross-network effects yields lower AIC values. (b) Between-type repetition effects  across  $1000$ simulations estimated using  a multiplex REM with individual effects in each layer, and a correctly specified non-separable model, with the same effect across layers. The red dashed line indicates the true value $\beta = 0.5$. Both approaches are unbiased but the non-separable model leads to more accurate estimates. }
    \label{fig:simulation_results}
\end{figure}

In order to test this hypothesis, we  use the Gillespie algorithm \citep{Gill77} to simulate a total of $2000$ events among $8$ actors from the multiplex relational event model~\eqref{eq:model_sim1}.  We use a nested case-control sampling scheme, where for each event a non-event of the same type is randomly sampled. Thus, both the relational event models of the misspecified case and those of the separable multiplex model are fitted via logistic regressions. 
Figure~\ref{fig:simulation_results}a reports the AIC of the models across $50$ simulation runs. The results show how a model based on within-type repetition does not adequately capture the cross-network effects of the underlying process, yielding systematically higher AIC values than the correctly specified multiplex relational event model. 

\subsection{Non-separable generating mechanism} \label{sec:sim_2}
In this second simulation, we evaluate the importance of accounting for model parameters that are shared across layers. As with the between-type covariates, this is a second dimension in which two processes can co-evolve. 
To this end, we consider a multiplex relational event model with two layers, $q$ and $q^*$, and intensity functions defined by
\begin{align*}
\lambda_{srq}(t) &= Y_{srq}(t) \lambda_0 \exp\left\{\beta \ x_{srq}^{(rep,q^*)}(t)\right\}\\
\lambda_{srq^*}(t) &= Y_{srq^*}(t) \lambda_0 \exp\left\{\beta \ x_{srq^*}^{(rep,q)}(t)\right\}.
\end{align*}
In particular, both the baseline hazard and the between-type repetition effect are shared across layers. We set these two parameters to  $\lambda_0 = 0.3$ and $\beta = 0.5$, respectively. The between-type repetition covariate measures the occurrence of past events between the same sender and receiver in the other layer. As with the previous simulation, we define this covariate  with an exponential temporal decay, namely
\[x_{srq}^{(rep,q^*)}(t) = \sum_{t^*_{srq^*} < t} e^{-0.5 (t - t^*_{srq^*})},\]
and similarly for $x_{srq^*}^{(rep,q)}(t)$ .
The fact that the between-type repetition effect $\beta$ is positive means that past events from sender \(s\) to receiver \(r\) of one type increase the instantaneous rate of an event from \(s\) to \(r\) of the other  type.

We simulate $n=200$ multiplex relational events from this model across $8$ actors  
 using the Gillespie algorithm \citep{Gill77}. For each event, we randomly sample a non-event from the same type. We repeat the simulation and the sampling $1000$ times. 
 For each simulated dataset, we compare two approaches. A first approach considers the correctly specified model. This is a multiplex relational event model with a between-type repetition effect in each layer and all parameters shared across layers. This can be fitted via a single logistic regression model, with a design matrix that collects information from events across all layers as discussed in section~\ref{sec:full_data}. In a second approach, we fit two separate relational event models, joined by a between-type repetition covariate at each layer. This second approach is the same used in the previous simulation and allows event-type specific parameters.  
Figure~\ref{fig:simulation_results}b visualizes the between-type repetition effects across the $1000$ simulations. The first two boxplots show the estimates under a separable model, one for each of the two layers, respectively, while the third boxplot shows the estimate of the shared parameter across the two layers using a non-separable model. The figure shows how both procedures provide unbiased estimates of the true parameter, but the non-separable model yields a more accurate estimate. Indeed, in this second case, the estimation benefits from the borrowing of information across the two layers. 

\section{Study of  gossiping dynamics via a multiplex REM} \label{sec:realdata}
We illustrate the proposed multiplex relational event framework using the school survey data discussed in section~\ref{sec:motivating_example}. As presented in that section, the data contain two directed relational event types.  
Specifically,  \textit{self-reported gossiping} captures students' reported gossip behaviour, whereas \textit{perceived gossiping} captures students' perceptions of being the target of gossip. We expect a complex underlying dynamics, whereby past events within each type as well as events in the other type affect the occurrence of future events across the two layers. 
Section~\ref{sec:gossip_model} presents a multiplex relational event model to study the co-evolution of self-reported and perceived gossiping. This is followed by section~\ref{sec:gossip_result}, which discusses the fitting of this model to the survey data and how the results inform us about the underlying dynamics of gossiping.

\subsection{Multiplex model formulation} \label{sec:gossip_model}
We denote the two gossiping types with $q_{\mathrm{self}}$ and $q_{\mathrm{perc}}$, respectively.
We define the following multiplex relational event model
\[
\lambda_{srq}(t)= Y_{srq}(t) \, \lambda_{0q}(t) \, \exp\left(f_q(\boldsymbol{x}_{srq}(t)) +  \boldsymbol{\gamma}^\top \boldsymbol{z}_{srq}\right), \quad q \in \{q_{\mathrm{self}}, q_{\mathrm{perc}}\},\]
where $\boldsymbol{x}_{srq}(t)$ denotes the vector of covariates for the tuple $(s,r)$ at time $t$ and layer $q$. 
 The function $f_q(\cdot)$ represents the overall contribution of the covariates to the intensity of gossiping of type $q$. The two models for each layer share a similar structure but differ in the specific covariates included in each model. Table~\ref{tab:biplex_effects} summarizes the fixed and random effects included in each model. For the fixed effects, these are split into exogenous and endogenous covariates, with the latter further split into within-type and between type covariates. A second column denotes whether these covariates enter the model linearly or through a smooth function.
 \begin{table}[t]
\centering
\begin{subtable}[t]{0.48\textwidth}
\centering
\caption{Self-reported gossip model}
\label{tab:effects_self}
\footnotesize
\setlength{\tabcolsep}{10pt}
\renewcommand{\arraystretch}{1.15}
\begin{tabularx}{\linewidth}{X l}
\hline
\textbf{Effect} & \textbf{Type} \\
\hline
\multicolumn{2}{l}{\textit{Exogenous}} \\
girl ego & linear   \\
girl alter & linear   \\
same gender & linear  \\
time & smooth \\
\hline
\multicolumn{2}{l}{\textit{Within-type endogenous (self $\to$ self)}} \\
repetition & smooth \\
reciprocity & smooth \\
sender out-degree & smooth \\
receiver in-degree & smooth \\
transitive closure & smooth \\
cyclic closure & smooth \\
sender balance & smooth \\
receiver balance & smooth \\
\hline
\multicolumn{2}{l}{\textit{Between-type endogenous (perceived $\to$ self)}} \\
immediate retaliation & linear \\
delayed retaliation & smooth \\
sender in-degree & smooth \\
receiver out-degree & smooth \\
\hline
\multicolumn{2}{l}{\textit{Random effects}} \\
class & random  \\
sender & random  \\
receiver & random  \\
\hline
\end{tabularx}
\end{subtable}
\hfill
\begin{subtable}[t]{0.48\textwidth}
\centering
\caption{Perceived gossip model}
\label{tab:effects_perceived}
\footnotesize
\setlength{\tabcolsep}{10pt}
\renewcommand{\arraystretch}{1.15}
\begin{tabularx}{\linewidth}{X l}
\hline
\textbf{Effect} & \textbf{Type} \\
\hline
\multicolumn{2}{l}{\textit{Exogenous}} \\
girl ego & linear  \\
girl alter & linear \\
same gender & linear  \\
time & smooth \\
\hline
\multicolumn{2}{l}{\textit{Within-type endogenous (perceived $\to$ perceived)}} \\
repetition & smooth  \\
reciprocity & smooth  \\
\hline
\multicolumn{2}{l}{\textit{Between-type endogenous (self $\to$ perceived)}} \\
immediate retaliation & linear \\
delayed retaliation & smooth \\
sender out-degree & smooth \\
receiver in-degree & smooth \\
sender in-degree & smooth \\
receiver out-degree & smooth \\
transitive closure & smooth \\
cyclic closure & smooth \\
sender balance & smooth \\
receiver balance & smooth \\
\hline
\multicolumn{2}{l}{\textit{Random effects}} \\
class &  random  \\
sender & random  \\
receiver & random  \\
\hline
\end{tabularx}
\end{subtable}
\caption{Overview of linear, smooth and random effects included in the multiplex relational event model for the co-evolution of (a) self-reported and (b) perceived gossiping. In each layer, the fixed effects are further split into exogenous and endogenous (within-type and between-type) covariates. A second column indicates whether these covariates enter the model linearly or via a smooth term.}
\label{tab:biplex_effects}
\end{table}
 We discuss below the definition and interpretation of the covariates that have been included in the model. 

In both layers, the models include  exogenous covariates that capture the effect of gender on gossiping. In particular, we consider the following three covariates:
\begin{align*}
x_{srq}^{(\mathrm{girl\_ego})} &= 1_{\{\mathrm{gender}(s)=\mathrm{female}\}},\\
x_{srq}^{(\mathrm{girl\_alter})} &= 1_{\{\mathrm{gender}(r)=\mathrm{female}\}},\\
x_{srq}^{(\mathrm{same\_gender})} &= 1_{\{\mathrm{gender}(s)=\mathrm{gender}(r)\}}.
\end{align*}
The first two capture differences in the propensity of female students to report or perceive gossip as senders and receivers, respectively, while the third variable captures whether interactions are more likely between students of the same gender, indicating a possible homophily effect. 

Within-type endogenous mechanisms capture how past activity within the same layer influences the occurrence of future events of the same type. To this end, both models include a selection of endogenous variables that are typically considered also for standard relational event models. These include repetition, reciprocity, degree-based effects, and higher-order structural effects, such as transitive closure, cyclic closure, and sender and receiver balance. For these variables, we use the definitions from Table~\ref{tab:covariates}. We choose which of these variables to include in each layer depending on whether it is about perceived or self-reported gossiping. A variable like transitive closure, for example, is relevant only in the context of actual gossiping behaviour and not at the level of perception.

The between-type covariates are those that are most relevant in the context of the proposed methodology. 
These variables allow past or contemporaneous activity in one event type to influence the occurrence of events in the other, thereby capturing cross-network dynamics between gossiping behaviour and its perception. We discuss the interpretation of these variables in more detail in the following. The definitions are taken from Table~\ref{tab:covariates}, or easily adapted from these. 

\paragraph{Between-type effects on self-reported gossip (perceived $\to$ self, Table~\ref{tab:biplex_effects}a).}
These effects capture how perceived gossiping influences reported gossiping behaviour. In particular, \textit{immediate retaliation} captures whether $s$ perceiving some gossip from $r$ is associated with an immediate response in the form of a self-reported gossip from $s$ towards $r$ within the same wave. In contrast to this, \textit{delayed retaliation} captures whether past perceived gossiping triggers a current self-reporting response. The degree effects capture whether the activity of students in the perceived gossip network has an effect on their self-reporting gossiping. In particular, \textit{sender in-degree} evaluates whether students who are frequently perceived as targets of gossip are more likely to report gossiping about others, whereas \textit{receiver out-degree} assesses whether students perceived as active gossipers are more likely to be targeted by reported gossip.

\paragraph{Between-type effects on perceived gossip (self $\to$ perceived, Table~\ref{tab:biplex_effects}b).}
These effects capture how self-reported gossiping shapes students' perceptions of being targeted. To this end, the \textit{immediate} and \textit{delayed retaliation} capture whether students perceive gossip from others as a response to their own current or past, respectively, reported gossiping towards those individuals. Among the degree effects, \textit{sender out-degree} assesses whether active gossipers are more likely to be perceived as gossipers, \textit{receiver in-degree} whether actual targets are more likely to perceive gossip, \textit{sender in-degree} whether students who are targeted by reported gossip are perceived as gossipers, and \textit{receiver out-degree} whether active gossipers are more likely to be perceived as targets. Among the higher-order effects, triadic closure captures whether $r$ is more likely to perceive gossiping from $s$ either when $s$ has gossiped about $a$ and $a$ about $r$ in the past (\textit{transitive closure}) or if  $r$ has gossiped about $a$ and $a$ about $s$ in the past (\textit{cyclic closure}). Similarly, triadic balance  captures whether $r$ is more likely to perceive gossiping from $s$ either when there is a third student $a$ gossiping about both $s$ and $r$ (\textit{sender balance}) or being gossiped about both by $s$ and $r$ (\textit{receiver balance}).

In addition to the exogenous and endogenous effects described above, the multiplex relational event model includes random effects at each layer.  In particular, we include random effects at the class, sender and receiver level. These capture specific gossiping activities within the same class or by same sender or the same receiver, which are not captured by the covariates included in the model, possibly due to classes' or students' characteristics which are not measured by the study. By including random effects, firstly we can perform the analysis from the full data, rather than separately for each classroom as in \cite{nynkegossip}. This leads to a more accurate estimation of the effects, by borrowing information across the different datasets. Secondly,
accounting for heterogeneity at the sender and receiver level reduces the risk of attributing persistent individual differences to endogenous network effects, and thus of detecting spurious structural effects \citep{rutaheterogeneity}.





\subsection{Dynamics of gossiping inferred from RECENS school survey} \label{sec:gossip_result}

We now fit the multiplex relational model described in section~\ref{sec:gossip_model} to the school survey data of section~\ref{sec:motivating_example}. As already explained in section~\ref{sec:motivating_example}, the data contain only censored information about the occurrence of events. In particular, we only have information about whether at least one event of each type occurred in each of the four waves, but no knowledge is available about the number of events and/or their exact timing. We therefore use the inferential procedure described in section~\ref{sec:interval_rightcensoring}. 
The four survey waves differ in duration, so we define the observation intervals in months as $(0,2]$, $(2,8]$, $(8,20]$, and $(20,32]$.
The covariates are evaluated for all the events in the risk set across the four waves. Sender and receiver degree are further transformed by taking the square-root, in order to improve computational stability. Since all covariates are measured only at the extreme of the time intervals, we approximate the integrals in the likelihood~\eqref{eq:fulllik_int_cens} using the average for each time interval. This is with the exception of repetition and degree-based effects. These are evaluated at the beginning of the time interval, since knowledge of these variables at the right extreme of the interval would provide information about the actual event that occurred.

We fit a Binomial generalized additive mixed model with a cloglog link separately for the self-reported and perceived gossiping events, respectively. That is, we fit a separable multiplex relational event model, with layer-specific effects for the variables that are in common across the two layers.  The length of the time intervals is incorporated through an offset which relates to the duration of the wave.  Non-linear effects are estimated using thin-plate regression splines through the \texttt{bam} function from the \texttt{mgcv} package \citep{wood2017}. Automatic model selection is performed using the double-penalty approach of \citet{penaltySelectTrue}, which allows uninformative smooth effects to be shrunk entirely to zero. Table~\ref{tab:self_net} reports the fitting of the self-reported gossiping component of the multiplex relational event model, while Table~\ref{tab:perc_net} is for the perceived gossiping component.  We discuss below the individual effects in each of the two components.

\begin{table}[t]{
\begin{center}
\begin{tabular*}{\textwidth}{@{\extracolsep{\fill}}lccccc}
\multicolumn{6}{c}{\textbf{estimated effects of the self-reported gossip model}}\\[2ex]
\textbf{fixed effects} & estimate & std. error & z value & p-value &  \\ 
  \hline
intercept & -8.618 & 0.330 & -26.080 & $<2\times10^{-16}$ & *** \\ 
  girl ego & 0.496 & 0.114 & 4.349 & 1.37e-05 & *** \\ 
  girl alter & 0.092 & 0.068 & 1.361 & 1.73e-01 &  \\ 
  same gender & 0.310 & 0.046 & 6.747 & 1.50e-11 & *** \\ 
  immediate retaliation (P) & 2.585 & 0.056 & 46.195 & $<2\times10^{-16}$ & *** \\
   \hline\\[0.1ex]
\textbf{smooth effects}   & edf & ref.df & chi.sq & p-value &  \\ 
  \hline
time & 1.858 & 3 & 179.570 & 3.02e-04 & *** \\ 
repetition (S) & 1.667 & 3 & 115.882 & $<2\times10^{-16}$ & *** \\ 
reciprocity (S) & 0.806 & 7 & 10.474 & 2.25e-02 & * \\ 
sender out-degree (S) & 4.919 & 9 & 5939.663 & $<2\times10^{-16}$ & *** \\ 
receiver in-degree (S) & 2.608 & 9 & 217.952 & $<2\times10^{-16}$ & *** \\ 
transitive closure (S) & 1.352 & 9 & 5.813 & 2.04e-01 &  \\ 
cyclic closure (S) & 0.794 & 9 & 3.635 & 1.06e-01 &  \\ 
sender balance (S) & 0.246 & 9 & 0.452 & 2.90e-01 &  \\ 
receiver balance (S) & 0.129 & 9 & 0.195 & 2.73e-01 &  \\ 
delayed retaliation (P) & 1.197 & 3 & 21.172 & 2.71e-03 & ** \\ 
receiver out-degree (P) & 0.155 & 9 & 0.655 & 2.71e-01 &  \\ 
sender in-degree (P) & 1.516 & 9 & 46.491 & 1.23e-02 & * \\
\hline \\[0.1ex]
\textbf{random effects} & \# groups & std. dev. & 95\% CI  & \\
\hline
class     & 44 & 0.516 & [0.376, 0.707] &  \\
receiver  & 1695 & 0.822 & [0.762, 0.887] &  \\
sender    & 1695 & 1.651 & [1.566, 1.741] &  \\
\hline
\end{tabular*}%
\end{center}}
\caption{Analysis of school survey data: estimates of fixed, smooth and random effects for the self-reported gossiping component of the multiplex relational event model. (S) denote the within-type endogenous covariates, while (P) denote the between-type endogenous covariates. For the fixed effects significance is assessed through the associated p-values, while the effective degrees of freedom (edf) indicate the complexity of the estimated non-linear relationships. \label{tab:self_net}}
\end{table}

\begin{figure}[t]
    \centering
    \begin{minipage}[b]{0.49\linewidth}
        \centering        \includegraphics[width=\linewidth]{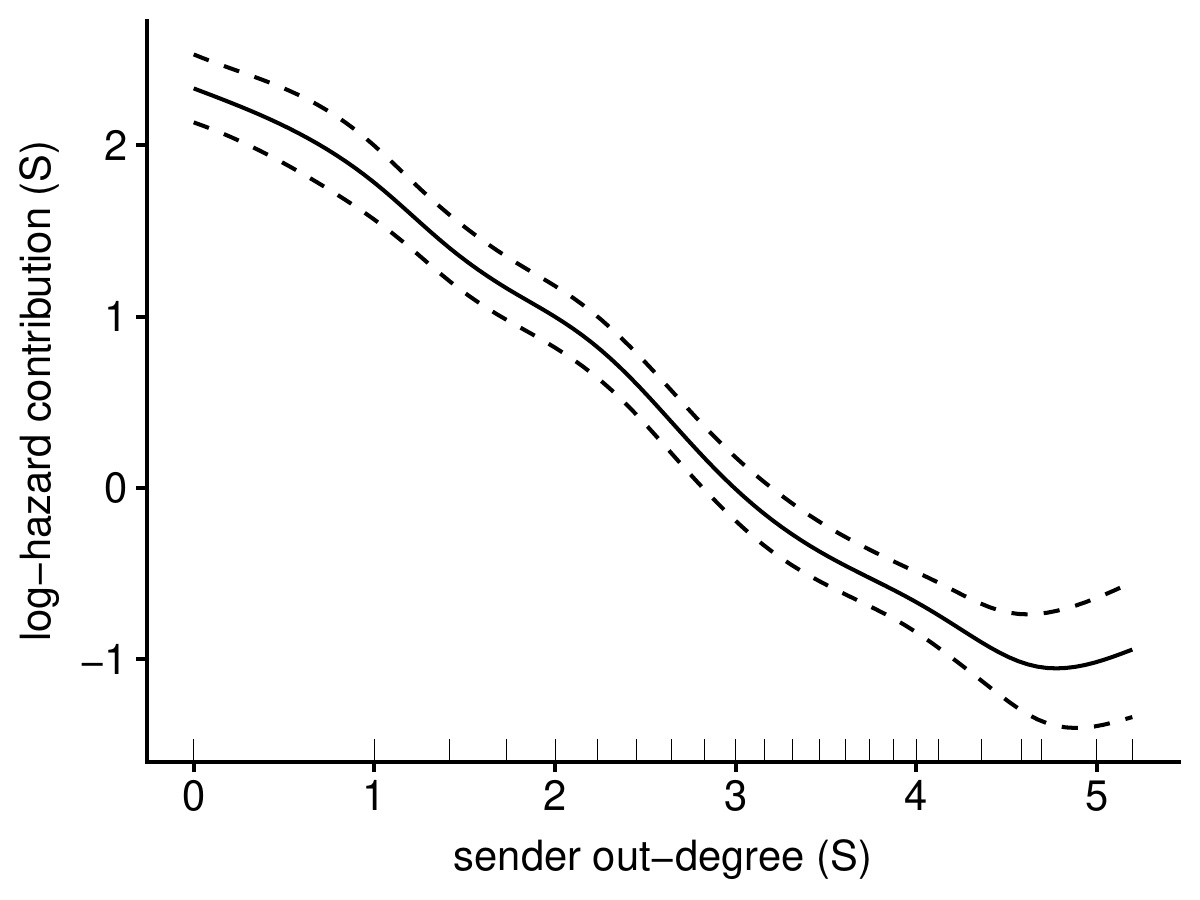}\\
(a)    
\end{minipage}
    \hfill
    \begin{minipage}[b]{0.49\linewidth}
        \centering
        \includegraphics[width=\linewidth]{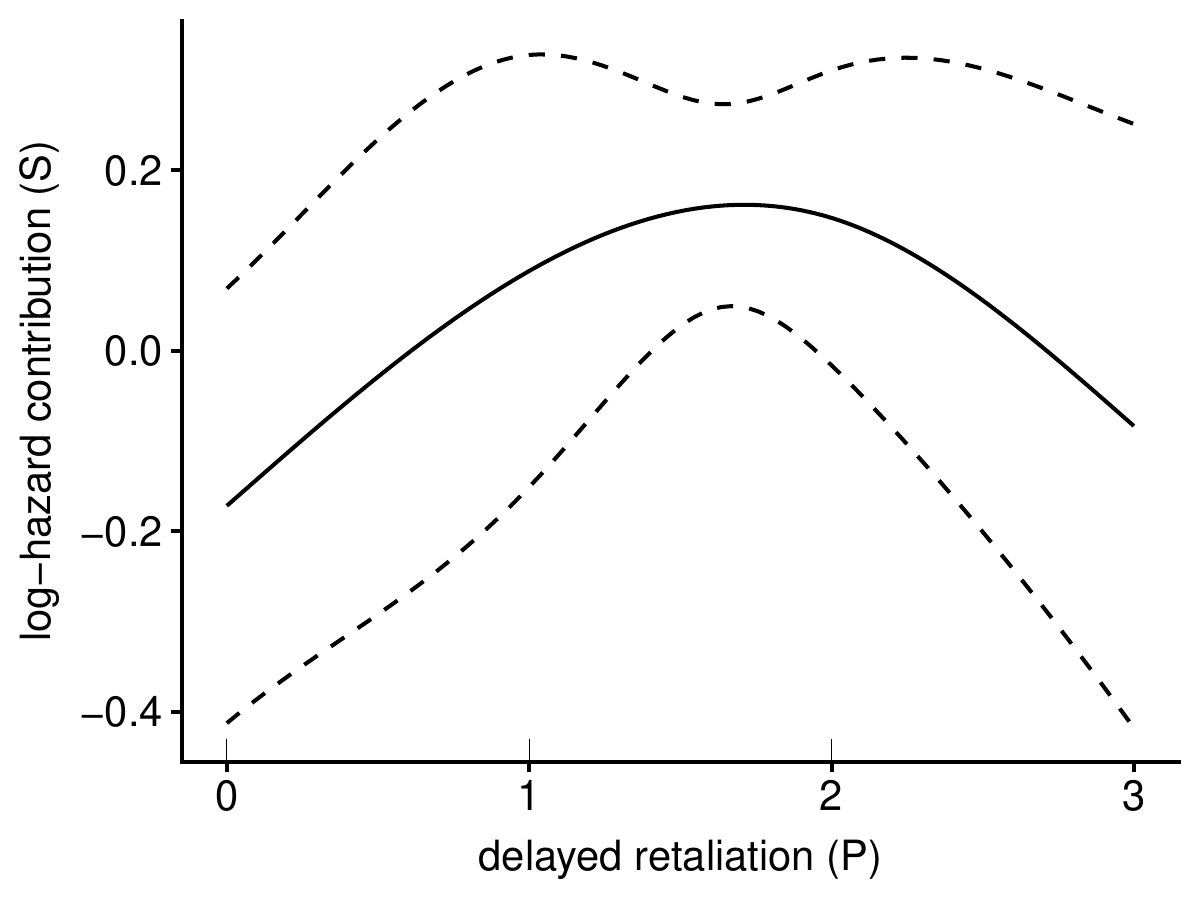}\\
(b)
    \end{minipage} 
        \begin{minipage}[b]{0.49\linewidth}
        \centering        \includegraphics[width=\linewidth]{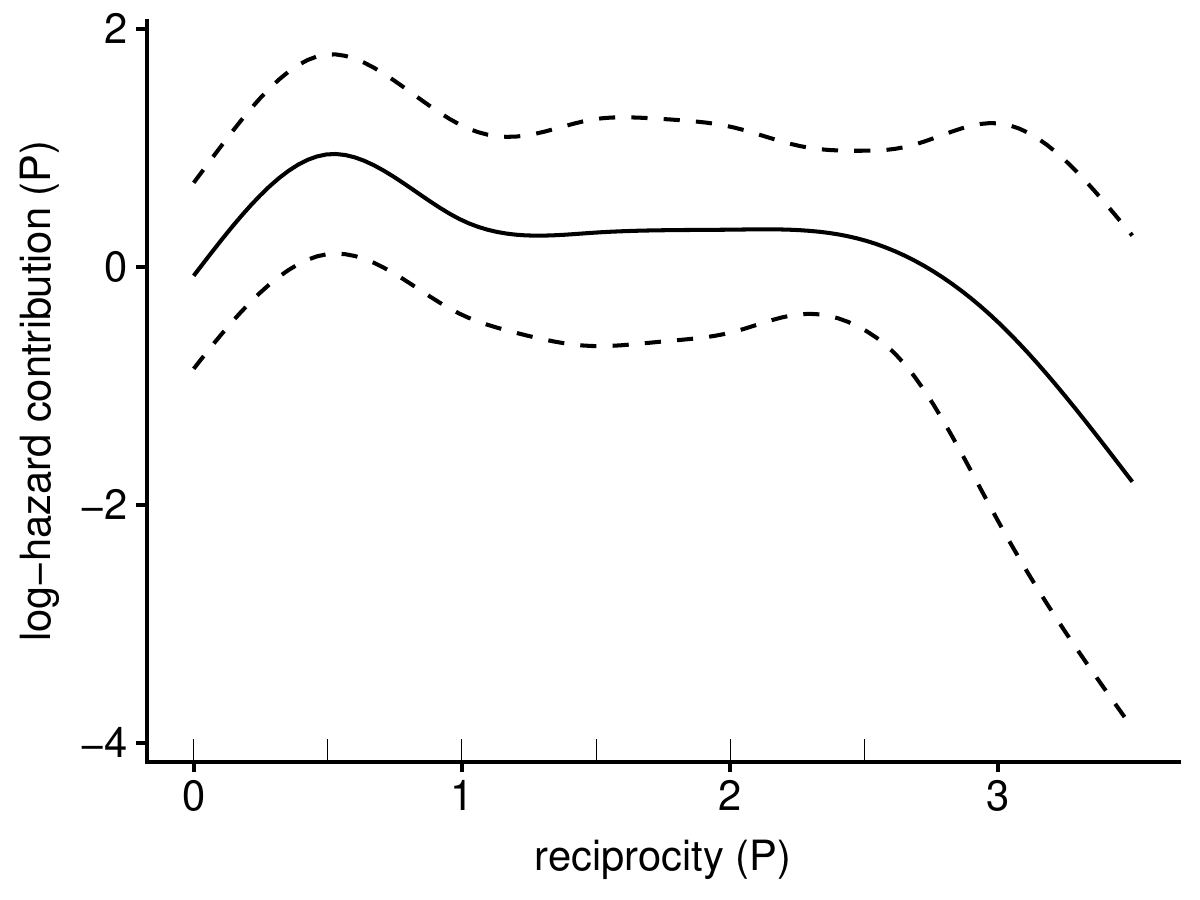}\\
(c)    
\end{minipage}
    \hfill
    \begin{minipage}[b]{0.49\linewidth}
        \centering
        \includegraphics[width=\linewidth]{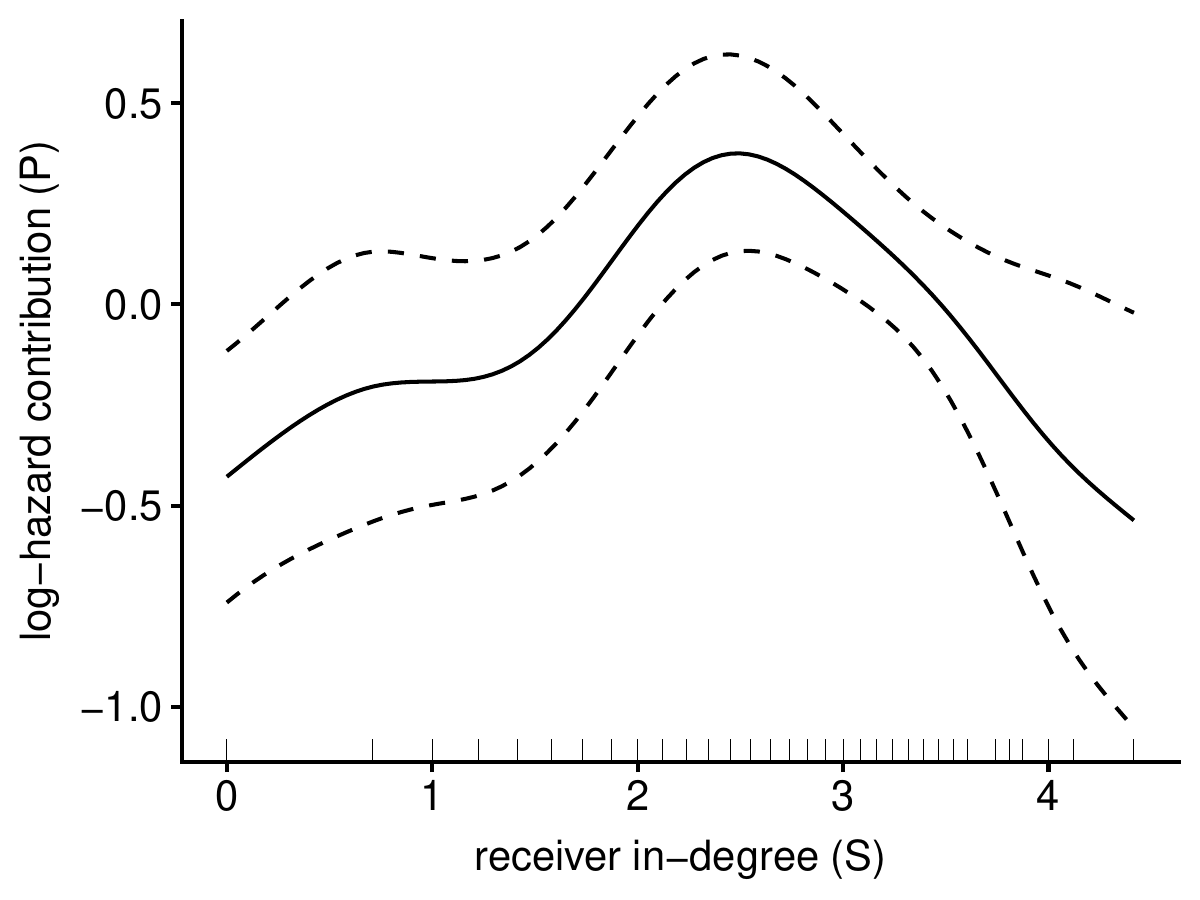}\\
(d)
    \end{minipage}  \hfill
    \caption{Estimated smooth effects for the multiplex relational event model of self-reported and perceived gossiping: (a) self-reported sender out-degree effect on self-reported gossiping (within-type effect) (b) delayed self-reported retaliation effect on perceived gossiping (between-type) (c) the effect of reciprocation on perceived gossiping (within-type) and (d) the effect of self-reported receiver activity on perceived gossiping (between-type). In all plots, the effect declines at higher levels, suggesting low levels of gossiping virality. The solid line represents the estimated effect, while the dashed lines indicate the 95\% confidence interval.}
    \label{fig:smooth-effects-gossiping}
\end{figure}

\paragraph{The dynamics of gossiping.}
The fixed effects in Table~\ref{tab:self_net} show a strong contribution of gender in self-reported gossiping. Female students are associated to a higher propensity to initiate gossiping than male students (\texttt{girl$\_$ego} positive and significant) and same-gender gossiping is approximately $36\%$ more likely than gossiping among students of different gender. This suggests a gender homophily effect in self-reported gossiping, as reported also by other studies \citep{nynkegossip, KisfalusiGossipReputation, poda2025hypereventnetworkmodellingpartially}.  Among the fixed effects, \texttt{immediate retaliation} is also highly significant. This is a between-type covariate. Its strong positive association indicates that students tend to respond promptly when they perceive themselves as targets, showing a co-evolution of perceived and self-reported gossiping.
Figure~\ref{fig:smooth-effects-gossiping} (top panel) reports the smooth effects for two of the significant endogenous covariates. The remaining non-linear effects for both networks are reported in the Supplementary Materials. The within-type sender out-degree effect (Figure~\ref{fig:smooth-effects-gossiping}a) shows a generally decreasing pattern suggesting that persistent self-reporting gossiping by the same individual is generally rare. This indicates a lack of virality in gossiping behaviour.
Combined with the significant standard deviation associated to the sender random effect, this shows how the inclusion of random effects in the model disentangles heterogeneity at the student level, which is driven primarily by unobserved students' traits. A similar behaviour is observed for the between-type delayed retaliation effect (Figure~\ref{fig:smooth-effects-gossiping}b), where the retaliatory self-reporting increases at low exposure, but flattens and declines at higher levels. One possible explanation for this specific pattern is that among the students reporting high levels of perceived gossip there are those who experience peer victimization or bullying. These students may be frequent targets of negative social attention but may also have fewer opportunities or less willingness to retaliate through gossiping, leading to a weaker association at higher levels of exposure.   

\begin{table}[t]{
\begin{center}
\begin{tabular*}{\textwidth}{@{\extracolsep{\fill}}lccccc}
\multicolumn{6}{c}{\textbf{estimated effects of the perceived gossip model}}\\[2ex]
\hline
\textbf{fixed effects} & estimate & std. error & z value & p-value &  \\ 
  \hline
intercept & -7.674 & 0.630 & -12.176 & $<2\times10^{-16}$ & *** \\ 
  girl ego & 0.108 & 0.106 & 1.022 & 3.07e-01 &  \\ 
  girl alter & 0.126 & 0.069 & 1.831 & 6.71e-02 & · \\ 
  same gender & 0.437 & 0.057 & 7.724 & 1.12e-14 & *** \\ 
  immediate retaliation (S) & 2.707 & 0.060 & 45.011 & $<2\times10^{-16}$ & *** \\ 
\hline \\[0.1ex]
\textbf{smooth effects} & edf & ref.df & chi.sq & p-value &  \\ 
  \hline
time & 2.968 & 3 & 1366.527 & $<2\times10^{-16}$ & *** \\ 
repetition (P) & 0.863 & 3 & 18.369 & 6.39e-03 & ** \\ 
reciprocity (P) & 3.656 & 7 & 156.765 & $<2\times10^{-16}$ & *** \\ 
delayed retaliation (S) & 1.210 & 3 & 11.941 & 4.47e-03 & ** \\ 
receiver in-degree (S) & 4.018 & 9 & 390.408 & $<2\times10^{-16}$ & *** \\ 
sender out-degree (S) & 1.188 & 9 & 33.338 & 3.05e-04 & *** \\ 
receiver out-degree (S) & 4.973 & 9 & 262.809 & $<2\times10^{-16}$ & *** \\ 
sender in-degree (S) & 2.353 & 9 & 86.213 & 1.01e-06 & *** \\ 
cyclic closure (S) & 1.335 & 9 & 8.202 & 2.77e-02 & * \\ 
transitive closure (S) & 0.639 & 9 & 3.555 & 9.21e-02 & · \\ 
sender balance (S) & 2.495 & 9 & 13.791 & 3.88e-02 & * \\ 
receiver balance (S) & 1.148 & 9 & 6.616 & 3.23e-02 & * \\  
   \hline\\[0.1ex]
\textbf{random effects} & \# groups & std. dev. & 95\% CI  & \\
\hline
class     & 44 & 0.441 & [0.311, 0.625] &  \\
receiver  & 1695 & 0.582 & [0.522, 0.649] &  \\
sender    & 1695 & 1.273 & [1.200, 1.350] &  \\
\hline
\end{tabular*}
\end{center}}
\caption{Analysis of school survey data: estimates of fixed, smooth and random effects for the perceived gossiping component of the multiplex relational event model. (P) denote the within-type endogenous covariates, while (S) denote the between-type endogenous covariates. For the fixed effects significance is assessed through the associated p-values, while the effective degrees of freedom (edf) indicate the complexity of the estimated non-linear relationships. \label{tab:perc_net}}
\end{table}
\paragraph{The dynamics of gossip perception.} 
The fixed effects in Table~\ref{tab:perc_net} show a prominent role played by gender homophily also on gossip perception, with students of the same gender approximately $55\%$ more likely to perceive gossiping between each other. Despite the same direction of the effect, the magnitude is significantly different compared to the same effect in the other layer. This is the case for all the other covariates that are in common across the two layers, supporting the use of a separable multiplex relational event model. 
The between-type \texttt{immediate retaliation} covariate is highly significant, suggesting a strong interdependence among the two processes. 
The positive estimate suggests that a student $r$ reporting gossiping about a student $s$ is more likely to perceive $s$ as gossiping about them.
Among the smooth effects, the within-type reciprocity effect (Figure~\ref{fig:smooth-effects-gossiping}c) shows a clear non-linear trend indicating that gossiping perceptions tend to become reciprocal at first, but their effect flattens out and eventually decreases at higher levels.
This indicates a lack of virality also in gossip perception. As before, this is accompanied by significant random effects at sender and receiver levels capturing the effect of unobserved students' traits on their perception of gossiping. 
 A similar pattern is visible for the between-type covariate receiver in-degree (Figure~\ref{fig:smooth-effects-gossiping}d), where an initial exposure to incoming self-reported gossip increases a student's propensity to perceive being targeted, but the effect levels off and declines at higher levels. 
 One possible explanation for this behaviour is that highly exposed individuals may become less sensitive to social cues or less likely to perceive additional gossip. As a result, they may remain less aware of these dynamics.


\section{Conclusion}

Gossiping unfolds across multiple and interconnected relational networks, with gossiping and gossip perception influencing each other over time. Understanding these cross-network effects is therefore essential for capturing the complex dynamics of gossiping. In this paper, we have presented a multiplex relational event model that includes both within-type and between-type effects, allowing us to study how events in one relational network influence, and are influenced by, events in another. We develop a general statistical inference framework for multiplex relational event models across general data structures. In all cases, we show how the fitting can be conducted efficiently via standard implementations of generalized additive models. Thus, models including linear, smooth and random effects across layers can be considered in order to study the dynamics of complex network phenomena. 

We illustrate the methodology on longitudinal survey data of Hungarian secondary schools, where two survey questions capture self-reported and perceived gossip, respectively. 
We fit a multiplex relational event model with a number of exogenous and endogenous covariates that capture the dynamics of gossiping and its co-evolution with the gossip perception process.
We include random effects at the class, sender and receiver levels. On the one hand, these account for a large heterogeneity among the classes. On the other hand, they allow us to disentangle a possible virality of gossiping from individual unobserved traits. The analysis reveals a number of highly significant between-type effects, showing the importance of accounting for cross-network dependences in order to have a more complete characterization of the dynamics of gossiping.

Beyond gossiping, the extension of relational event models to multiplex settings proposed in this paper provides a flexible framework for studying dynamic interactions across multiple interconnected relational networks, with potential applications across a variety of domains.

\section{Code availability}

The code for reproducing the simulation study and the empirical analysis of gossip data can be found at the \texttt{GitHub} repository page https://github.com/veronicapoda/multiplex-rems.git.

\section*{Acknowledgments}

We thank the Research Center for Educational and Network Studies (RECENS) for providing access to the data from the school survey and as well as providing further information. EW gratefully acknowledges funding from the Swiss National Science Foundation (grant number 240065).

\bibliographystyle{chicago}  
\bibliography{bib.bib}

\begin{thebibliography}{}

\bibitem[\protect\citeauthoryear{Bianchi, Filippi-Mazzola, Lomi, and
  Wit}{Bianchi et~al.}{2024}]{bianchi2023}
Bianchi, F., E.~Filippi-Mazzola, A.~Lomi, and E.~C. Wit (2024).
\newblock Relational event modeling.
\newblock {\em Annual Review of Statistics and Its Application\/}~{\em 11},
  297--319.

\bibitem[\protect\citeauthoryear{Borgan, Goldstein, and Langholz}{Borgan
  et~al.}{1995}]{Borgan1995Methods}
Borgan, {\O}., L.~Goldstein, and B.~Langholz (1995).
\newblock Methods for the analysis of sampled cohort data in the cox
  proportional hazards model.
\newblock {\em The Annals of Statistics\/}~{\em 23\/}(5), 1749--1778.

\bibitem[\protect\citeauthoryear{Boschi, Juozaitienė, and Wit}{Boschi
  et~al.}{2025}]{boschi_smooth_2025}
Boschi, M., R.~Juozaitienė, and E.~C. Wit (2025).
\newblock Mixed additive modelling of global alien species co-invasions of
  plants and insects.
\newblock {\em Journal of the Royal Statistical Society Series C: Applied
  Statistics\/}, qlaf034.

\bibitem[\protect\citeauthoryear{Boschi and Wit}{Boschi and
  Wit}{2026}]{boschi_introREM_2026}
Boschi, M. and E.~C. Wit (2026).
\newblock Introduction to relational event modelling.
\newblock arXiv:2604.07063.

\bibitem[\protect\citeauthoryear{Chen}{Chen}{2021}]{chen2021statistical}
Chen, T.-H.~Y. (2021).
\newblock Statistical inference for multilayer networks in political science.
\newblock {\em Political Science Research and Methods\/}~{\em 9\/}(2),
  380--397.

\bibitem[\protect\citeauthoryear{Cox}{Cox}{1975}]{cox_1975}
Cox, D.~R. (1975, 08).
\newblock Partial likelihood.
\newblock {\em Biometrika\/}~{\em 62\/}(2), 269--276.

\bibitem[\protect\citeauthoryear{Espelage, Green, and Wasserman}{Espelage
  et~al.}{2007}]{bullying}
Espelage, D., H.~Green, and S.~Wasserman (2007, 02).
\newblock Statistical analysis of friendship patterns and bullying behaviors
  among youth.
\newblock {\em New directions for child and adolescent development\/}~{\em
  118}, 61--75.

\bibitem[\protect\citeauthoryear{{Filippi-Mazzola} and Wit}{{Filippi-Mazzola}
  and Wit}{2024}]{filippi-mazzola2024}
{Filippi-Mazzola}, E. and E.~C. Wit (2024).
\newblock A stochastic gradient relational event additive model for modelling
  {{US}} patent citations from 1976 to 2022.
\newblock {\em Journal of the Royal Statistical Society Series C: Applied
  Statistics\/}~{\em 73\/}(4), 1008--1024.

\bibitem[\protect\citeauthoryear{Gillespie}{Gillespie}{1977}]{Gill77}
Gillespie, D.~T. (1977).
\newblock Exact stochastic simulation of coupled chemical reactions.
\newblock {\em The Journal of Physical Chemistry\/}~{\em 81\/}(25), 2340--2361.

\bibitem[\protect\citeauthoryear{Hong and Niezink}{Hong and
  Niezink}{2026}]{nynkegossip}
Hong, A. and N.~Niezink (2026).
\newblock The multiplex $p_2$ model: Mixed-effects modeling for multiplex
  social networks.
\newblock {\em Bayesian Analysis\/}~{\em 21}, 1527--1557.

\bibitem[\protect\citeauthoryear{Juozaitienė and Wit}{Juozaitienė and
  Wit}{2024a}]{juozaitiene_its_2024}
Juozaitienė, R. and E.~C. Wit (2024a).
\newblock It’s about time: revisiting reciprocity and triadicity in
  relational event analysis.
\newblock {\em Journal of the Royal Statistical Society Series A: Statistics in
  Society\/}~{\em 89\/}(1), qnae132.

\bibitem[\protect\citeauthoryear{Juozaitienė and Wit}{Juozaitienė and
  Wit}{2024b}]{rutaheterogeneity}
Juozaitienė, R. and E.~C. Wit (2024b).
\newblock Nodal heterogeneity can induce ghost triadic effects in relational
  event models.
\newblock {\em Psychometrika\/}~{\em 89\/}(1), 151--171.

\bibitem[\protect\citeauthoryear{Kisfalusi}{Kisfalusi}{2018}]{bullies}
Kisfalusi, D. (2018).
\newblock Bullies and victims in primary schools: The associations between
  bullying, victimization, and students’ ethnicity and academic achievement.
\newblock {\em Intersections\/}~{\em 4}, 133--158.

\bibitem[\protect\citeauthoryear{Kisfalusi, Tak{\'a}cs, and P{\'a}l}{Kisfalusi
  et~al.}{2019}]{KisfalusiGossipReputation}
Kisfalusi, D., K.~Tak{\'a}cs, and J.~P{\'a}l (2019).
\newblock Gossip and reputation in adolescent networks.
\newblock {\em The Oxford Handbook of Gossip and Reputation\/}.

\bibitem[\protect\citeauthoryear{Lembo, Juozaitien{\.e}, Vinciotti, and
  Wit}{Lembo et~al.}{2026}]{lembo_relational_2025}
Lembo, M., R.~Juozaitien{\.e}, V.~Vinciotti, and E.~C. Wit (2026).
\newblock Relational event models with global covariates: an application to
  bike sharing.
\newblock {\em Journal of the Royal Statistical Society Series C: Applied
  Statistics\/}~{\em 75\/}(3), 583--600.

\bibitem[\protect\citeauthoryear{Marra and Wood}{Marra and
  Wood}{2011}]{penaltySelectTrue}
Marra, G. and S.~N. Wood (2011).
\newblock Practical variable selection for generalized additive models.
\newblock {\em Computational Statistics and Data Analysis\/}~{\em 55\/}(7),
  2372--2387.

\bibitem[\protect\citeauthoryear{Meyer}{Meyer}{1962}]{meyer1962decomposition}
Meyer, P. (1962).
\newblock A decomposition theorem for supermartingales.
\newblock {\em Ill. J. Math.\/}~{\em 6\/}(2), 193--205.

\bibitem[\protect\citeauthoryear{Poda, Vinciotti, and Wit}{Poda
  et~al.}{2026}]{poda2025hypereventnetworkmodellingpartially}
Poda, V., V.~Vinciotti, and E.~C. Wit (2026).
\newblock Hyperevent network modelling of partially observed gossip data.
\newblock {\em Journal of the Royal Statistical Society Series C: Applied
  Statistics\/}, qlag041.

\bibitem[\protect\citeauthoryear{Snijders, Lomi, and Torl{\'o}}{Snijders
  et~al.}{2013}]{snijders2013model}
Snijders, T. A.~B., A.~Lomi, and V.~J. Torl{\'o} (2013).
\newblock A model for the multiplex dynamics of two-mode and one-mode networks,
  with an application to employment preference, friendship, and advice.
\newblock {\em Social Networks\/}~{\em 35\/}(2), 265--276.

\bibitem[\protect\citeauthoryear{Vu, Pattison, and Robins}{Vu
  et~al.}{2015}]{VU2015121}
Vu, D., P.~Pattison, and G.~Robins (2015).
\newblock Relational event models for social learning in moocs.
\newblock {\em Social Networks\/}~{\em 43}, 121--135.

\bibitem[\protect\citeauthoryear{Vörös, Boda, Csaba, Kisfalusi, Radó, Varga,
  and Takács}{Vörös et~al.}{2025}]{gossip_dataset}
Vörös, A., Z.~Boda, Z.~L. Csaba, D.~Kisfalusi, M.~Radó, K.~Varga, and
  K.~Takács (2025).
\newblock Studying school communities as multiplex dynamic networks: The
  {“RECENS Wired into Each Other”} dataset, 2010–2013.
\newblock {\em CONNECTIONS\/}.

\bibitem[\protect\citeauthoryear{Wang}{Wang}{2013}]{wang2013ergmextensions}
Wang, P. (2013).
\newblock Exponential random graph model extensions: Models for multiple
  networks and bipartite networks.
\newblock In D.~Lusher, J.~Koskinen, and G.~Robins (Eds.), {\em Exponential
  Random Graph Models for Social Networks}, Chapter~10. Cambridge University
  Press.

\bibitem[\protect\citeauthoryear{Wood}{Wood}{2017}]{wood2017}
Wood, S. (2017).
\newblock {\em Generalized Additive Models: An Introduction With R, Second
  Edition}.
\newblock CRC Press.

\end{thebibliography}


\appendix
\section*{Supplementary Materials}

\section{Additional non-linear effects for self-reported gossip}

\begin{figure}[t]
    \centering    \includegraphics[width=\textwidth]{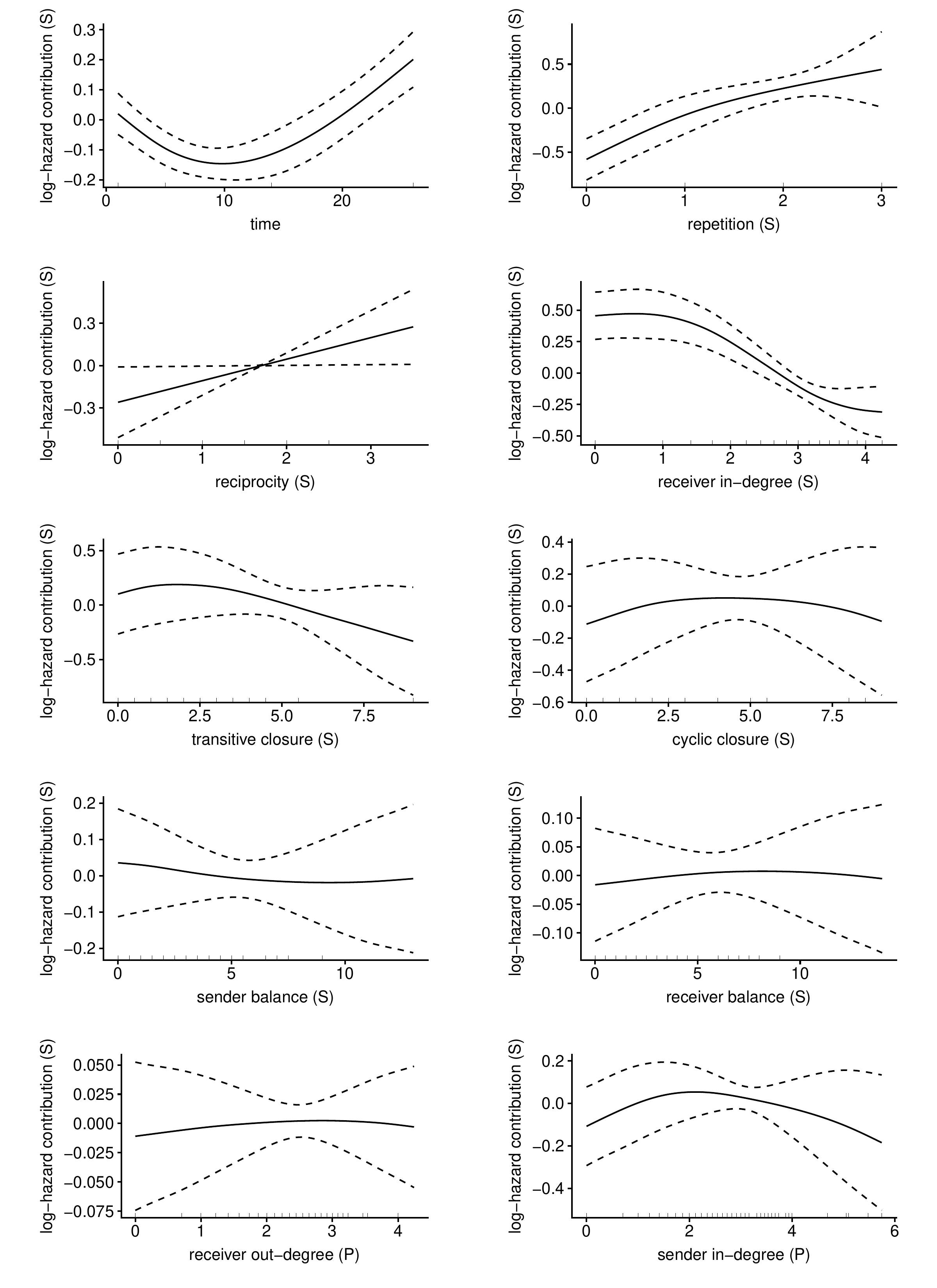}
    \caption{Estimated smooth effects for the multiplex relational event model of self-reported gossiping. The solid line represents the estimated effect, while the dashed lines indicate the 95\% confidence interval.}
    \label{fig:supp_model1}
\end{figure}

Figure~\ref{fig:supp_model1} shows the remaining non-linear effects for the self-reported gossip model. The time effect increases after 10 months, suggesting that students report more gossiping as time goes on, possibly because they become more confident and get to know each other better over the years of high school. Regarding the dyadic effects, both within-type repetition and reciprocity increase, showing that past self-reported gossiping increases the rate of future ones. In contrast, the within-type receiver in-degree decreases at higher values, which again confirms a low level of virality in gossiping, as also discussed in the main text. Triadic effects, such as transitive and cyclic closure, sender and receiver balance, are generally flat and not significant. Finally, about the  cross-layer dynamics, two between-type nodal effects, specifically receiver out-degree and sender in-degree, slightly increase and then plateau, once more confirming low virality in gossip dynamics.

\section{Additional non-linear effects for perceived gossip}

\begin{figure}[t]
    \centering    \includegraphics[width=\textwidth]{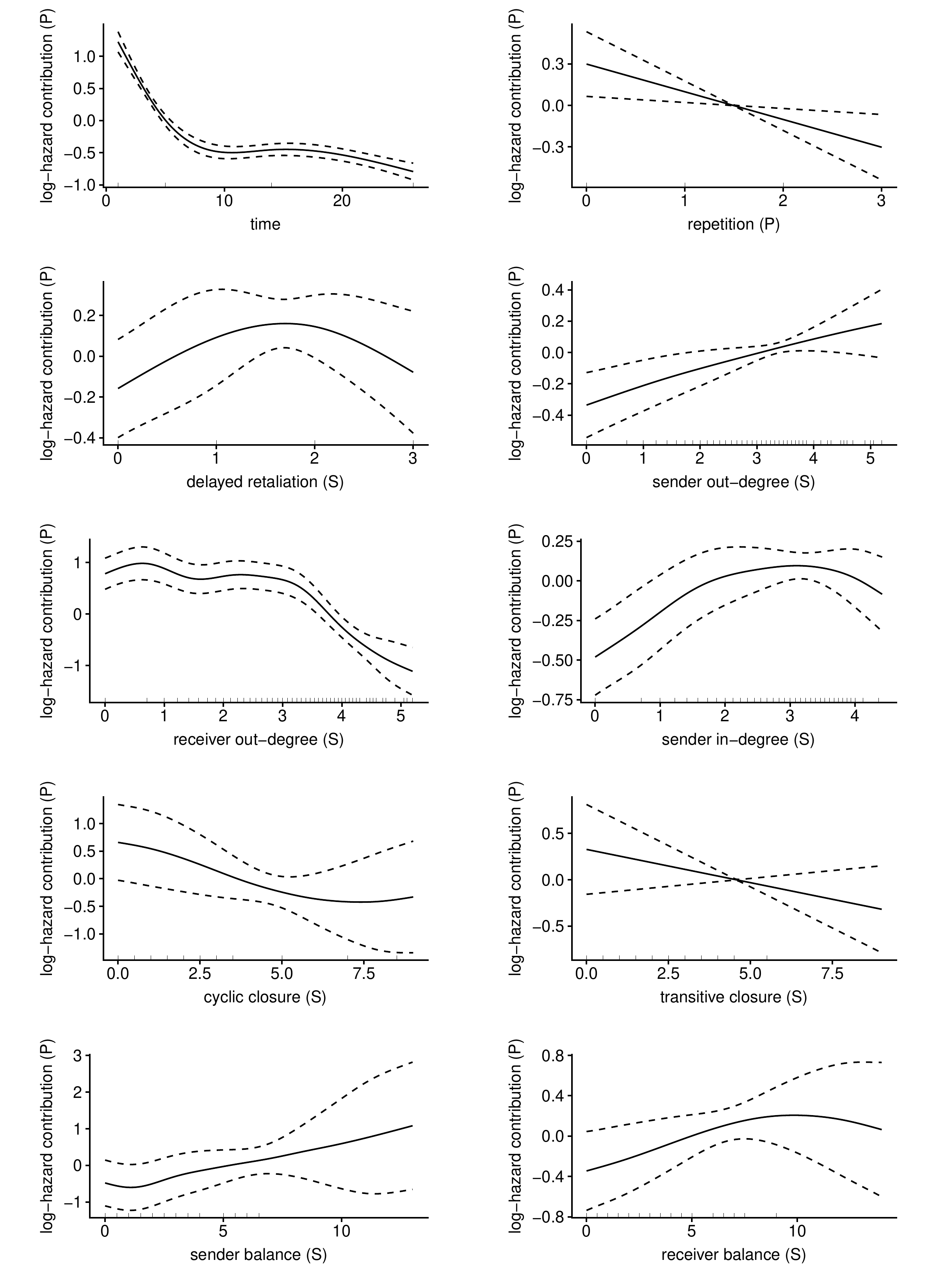}
    \caption{Estimated smooth effects for the multiplex relational event model of perceived gossiping. The solid line represents the estimated effect, while the dashed lines indicate the 95\% confidence interval.}
    \label{fig:supp_model2}
\end{figure}

Figure~\ref{fig:supp_model2} reports the remaining non-linear effects for the perceived gossip model. The curve of time is similar to that of self-reported gossip up to month $10$, after which the two trajectories diverge. Overall, the time effect is stronger in the perceived network, with a log-hazard contribution ranging from approximately -1 to 1, compared with approximately -0.2 to 0.2 in the self-reported network. In contrast to the self-reported gossip, here the dyadic effects, show a different trend, with a decreasing within-type repetition and a between-type delayed retaliation which slightly increases and then, after plateauing, decreases. Low gossip virality is once again confirmed by the between-type nodal effects, such as receiver out-degree and sender in-degree. Instead, there is some indication that students who gossip a lot in the self-reported network increase the rate of being perceived more, as shown by the increasing between-type sender out-degree, even though some uncertainty can be observed at higher values. Finally, between-type triadic effects like cyclic and transitive closure show a decreasing trend, while sender and receiver balance increase, at least initially.

\end{document}